\documentclass[review]{elsarticle}
\usepackage{graphicx}
\usepackage{subcaption}
\usepackage{color}
\journal{Journal of \LaTeX\ Templates}

\biboptions{authoryear}

\begin{document}
\newcommand{\lsimeq}{\mbox{$\, \stackrel{\scriptstyle <}{\scriptstyle
\sim}\,$}}
\newcommand{\gsimeq}{\mbox{$\, \stackrel{\scriptstyle >}{\scriptstyle
\sim}\,$}}
\newcommand*\txt[1]{_{\textnormal{#1}}}
\begin{frontmatter}

\title{Magnetic fields in isolated and interacting white dwarfs}


\author{Lilia Ferrario\footnote{Lilia.Ferrario@anu.edu.au} and Dayal Wickramasinghe\footnote{Dayal.Wickramasinghe@anu.edu.au}}
\address{Mathematical Sciences Institute, The Australian National
University, Canberra, ACT 2601, Australia}
\author{Adela Kawka\footnote{Adela.Kawka@curtin.edu.au}}
\address{International Centre for Radio Astronomy Research, Curtin University, Perth, WA 6102,
Australia}

\begin{abstract}

  The magnetic white dwarfs (MWDs) are found either isolated 
 or in interacting binaries. The isolated MWDs divide into two groups: a high field group ($10^5 - 10^9$\,G) comprising some $13 \pm 4$\% of all white dwarfs (WDs), and a low field group ($B<10^5$\,G) whose incidence is currently under investigation. The situation may be similar in magnetic binaries because the bright accretion discs in low field systems hide the photosphere of their WDs thus preventing the study of their magnetic fields' strength and structure. Considerable research has been devoted to the vexed question on the origin of magnetic fields. One hypothesis is that WD magnetic fields are of fossil origin, that is, their progenitors are the magnetic main-sequence Ap/Bp stars and magnetic flux is conserved during their evolution. The other hypothesis is that magnetic fields arise from binary interaction, through differential rotation, during common envelope evolution. If the two stars merge the end product is a single high-field MWD. If close binaries survive and the primary develops a strong field, they may later evolve into the magnetic cataclysmic variables (MCVs). The recently discovered population of hot, carbon-rich WDs exhibiting an incidence of magnetism of up to about 70\% and a variability from a few minutes to a couple of days may support the merging binary hypothesis. The fields in the weakly magnetic WDs may instead arise from a dynamo mechanism taking place in convective zones during post main-sequence evolution. Should this be the case, there may be a field strength below which all WDs are magnetic and thus fields are expected to always play a role in accretion processes in close binaries. Several studies have raised the possibility of the detection of planets around MWDs. Rocky planets may be discovered by the detection of anomalous atmospheric heating of the MWD when the unipolar inductor mechanism operates whilst large gaseous planets may reveal themselves through cyclotron emission from wind-driven accretion onto the MWD. Planetary remains have recently revealed themselves in the atmospheres of about 25\% of WDs that are polluted by elements such as Ca, Si, and often also Mg, Fe, Na. This pollution has been explained by ongoing accretion of planetary debris. Interestingly, the incidence of magnetism is approximately 50\% in cool, hydrogen-rich, polluted WDs, suggesting that these fields may be related to differential rotation induced by some super-Jupiter bodies that have plunged into the WD. The study of isolated and accreting MWDs is likely to continue to yield exciting discoveries for many years to come.
\end{abstract}


\end{frontmatter}

\section{Introduction}
\label{intro}

In recent years, there has been a rapid increase in the rate of discovery of MWDs. There are now around 600 isolated MWDs with magnetic fields spanning the range $10^{3}$\,G to $10^{9}$\,G and about 200 MWDs in interacting binaries. The sample is now large enough to allow studies of the group properties of the MWDs.

Early estimates of the space density of isolated MWDs were mostly based on those discovered by the Palomar Green (PG) survey which suggested an incidence of magnetism of only $\sim 2$\% for stars with fields greater than a few $10^6$\,G.  It was subsequently realised that this estimate should be revised upwards to at least 10\% to account for the different volumes that are sampled by WDs with different masses when allowance is made for the higher mean mass of the MWDs \citep{Liebert2003}. A higher incidence of magnetism was also indicated from studies of the nearly complete 13\,pc ($21\pm 8$\%) and 20\,pc ($13\pm 4$\%) volume limited samples of WDs conducted by \citet{Kawka2007}. These findings were more recently confirmed by \citet{Holberg2016} who found that the inner 20\,pc volume contains 12\% MWDs which translates to a space density of $\sim 0.6\times10^{-3}$\,pc$^{-3}$.

Spectropolarimetric surveys have revealed the existence of low-field MWDs with fields in the range $10^3- 10^5$\,G \citep[e.g.,][]{Aznar2004, Jordan2007, Landstreet2012}).  Their incidence is estimated by \citet{Landstreet2012} to be around 10\%, but this  field regime still needs to be thoroughly investigated.

The origin of the magnetic fields of isolated and binary WDs has received renewed attention in recent years. The traditional view has been that their progenitors are the magnetic main-sequence Ap and Bp stars, and that the fields are essentially of fossil origin. A compelling argument in favour of this hypothesis is the correspondence that exists between the field strengths of the two groups under magnetic flux conservation \citep{Ferrario2005a}. However, given the incidence of magnetism among WDs the required birth rate of MWDs exceeds by a factor $\sim 2 - 3$ the birth rate of Ap and Bp stars \citep{Kawka2004}. This and the lack of binaries consisting of a MWD paired with a fully detached, non-degenerate companion \citep{Liebert2005, Liebert2015, Tovmassian2016, Tovmassian2018a, Tovmassian2018b} has led to the suggestion that magnetic fields in WDs could be the outcome of binary interaction and stellar merging (see \S\,\ref{origin}).

It has been known for some time that MWDs tend to exhibit complex field structures and that off-centred dipole models generally produce a better fit to spectroscopic observations than centred dipole models \citep[see][]{Wickramasinghe2000}. Structures with dominant quadrupolar components or evidence for significant contributions from higher order multipoles have also emerged from detailed modelling of isolated and accreting MWDs \citep{Maxted2000, Euchner2005, Euchner2006, Beuermann2007, Landstreet2017}, although not all MWDs require complex field structures to understand their spectra.  This suggests that there may be a complex interplay between fossil and dynamo generated fields during stellar evolution and/or stellar merging events which are still poorly understood phenomena despite some extensive studies conducted over many decades \citep{Tout2004, Brun2005, Featherstone2009, Potter2010, Quentin2018}.

We begin by presenting an overview of the methods used for determining magnetic fields in WDs in \S\,\ref{measure}. In this section we also present the rich variety of spectra that result from the Zeeman effect in WDs of different atmospheric compositions.  The assumptions involved in the construction of model atmospheres for MWDs are presented in \S\,\ref{atmo} where we also describe the methods that have been used to unravel the magnetic field structure of MWDs. The population properties of MWDs (fields, mass, temperature distributions) are highlighted in \S\,\ref{pops}. The characteristics of interacting magnetic binaries can be found in \S\,\ref{binary}. The population properties of isolated and accreting MWDs and current ideas on the origin of magnetic fields are presented and discussed in \S\,\ref{origin}. A summary and an outlook for future research is presented in \S\,\ref{conclusions}.

\section{Measuring magnetic fields in WDs}\label{measure}

\subsection{Zeeman spectroscopy}\label{zeeman}

\subsubsection{Hydrogen lines}\label{Hlines}

Similarly to their non-magnetic counterparts, the vast majority of MWDs are hydrogen-rich and are classified as DAH, if the WD was originally discovered to be magnetic from Zeeman splitting, or DAP, if the WD was discovered to be magnetic from polarisation studies.  An understanding of their properties relies heavily on the theory of the hydrogen atom at strong magnetic fields.

The structure of the hydrogen atom at arbitrary field strengths was solved in the mid 1980's by various groups \citep{Kemic1974, Rosner1984, Forster1984, Henry1985, Wunner1985} culminating in the publication of wavelengths and oscillator strengths of most of the important transitions in the ultraviolet, optical and infrared spectral regions as a function of the parameter
\begin{equation}
\beta= {\omega_C \over 4\omega_R}= \frac{B}{4.7\times 10^9\,{\rm G}}
\end{equation}
that measures the relative importance of the cyclotron frequency of a free electron
\begin{equation}
\omega_{C}= \displaystyle{\frac{eB}{m_e c}}
\end{equation}
to the Rydberg frequency $\omega_{R}$. Here $e$ is the electron charge, $m_e$ the electron rest mass and $c$ the speed of light. At low $\beta$, the atomic structure is dominated by the spherical symmetry of the Coulomb potential while at high $\beta$ is dominated by the cylindrical symmetry of the field, $B$.

The dependence of wavelength $\lambda$ on magnetic field, the so-called $\lambda-B$ curves, have formed the basis for all studies of MWDs. These calculations are for fields that are high enough for the magnetic splitting to be larger than the spin-orbit coupling (the Paschen-Back regime). 

If we use the standard zero field quantum numbers $(n,l,m_l)$ to label energy levels  ($n$ is the principal quantum number, $l$ is the angular momentum quantum number, and $m_l$ is the magnetic quantum number), three basic Zeeman regimes can be identified: linear (removal of $m_l$ degeneracy), quadratic (removal of $l$ degeneracy), fully mixed (removal of $n$ degeneracy).

\begin{figure}[htbp]
\begin{center}
\includegraphics[width=10cm]{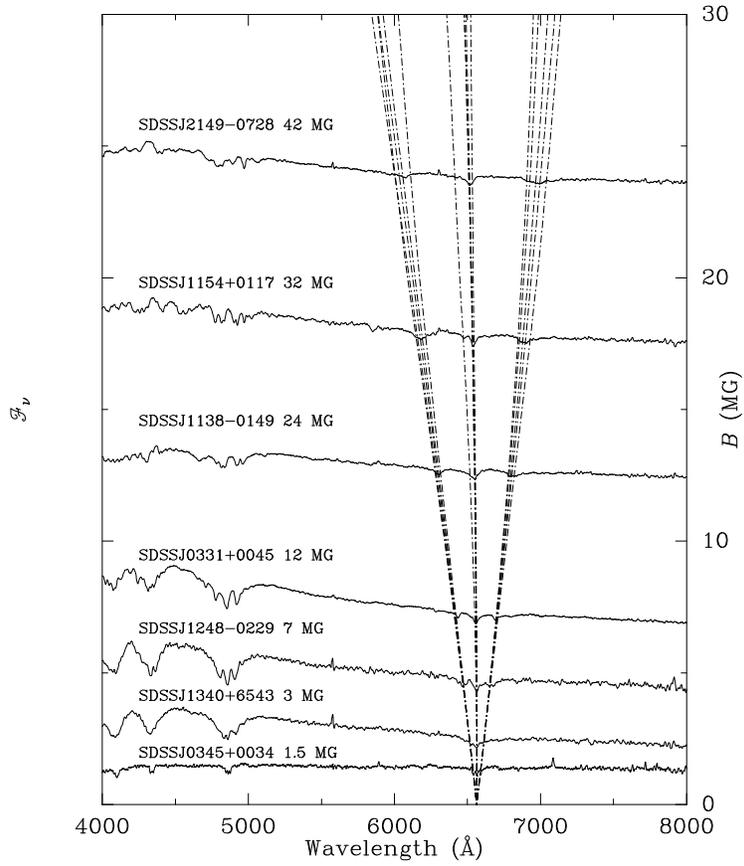} 
\caption{The Zeeman effect in the linear and quadratic regimes illustrated in the spectra of seven DAP WDs. The H$_\alpha$ line is split into a triplet over the entire field range($1.5-42$\,MG) while the quadratic Zeeman effect is already seen at H$_{\beta}$ at a field of 12\,MG \citep[from][]{Schmidt2003}.
}
\label{zeem_triplets_low}
\end{center}
\end{figure}

\begin{figure}[htbp]
\begin{center}
\includegraphics[width=8cm]{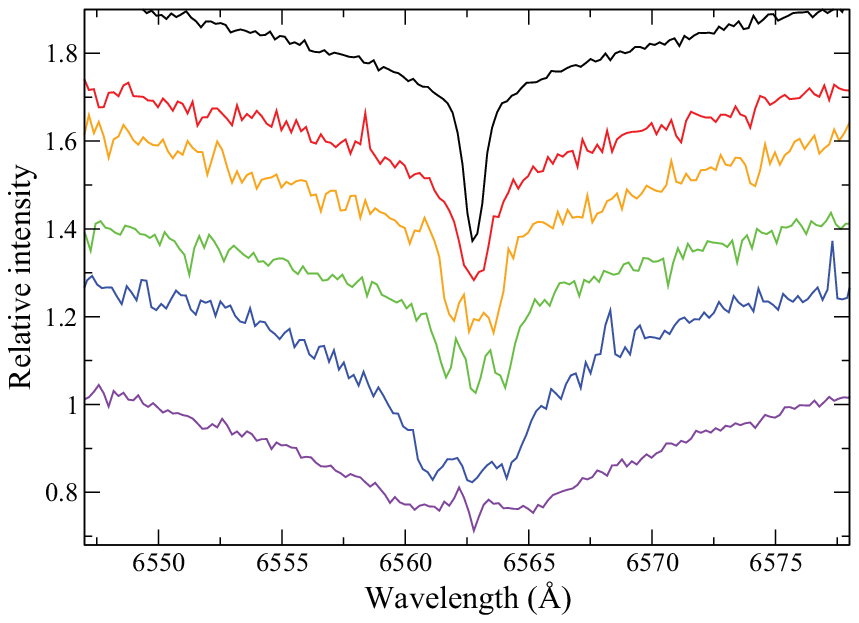}
\caption{Intensity spectrum of a non-magnetic WD (top). The spectra that follow are those of MWDs with increasingly higher field strength. For full details see \citet{Landstreet2016}.}
\label{Landstreet2016_Weak_Series}
\end{center}
\end{figure}

\begin{figure}[htbp]
\begin{center}
\includegraphics[width=10cm]{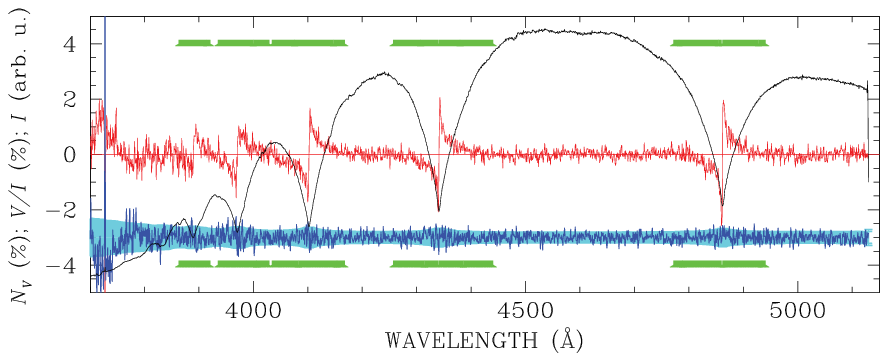}
\caption{Spectropolarimetric observations of WD\,1105$-$340. The black curve is the intensity profile and the red curve the circular polarisation (V/I) spectrum in percentage units \citep[see][for full details]{Landstreet2019}. The change in the sign of circular polarisation across the members of the Balmer series is clearly visible.}
\label{Landstreet_LowField}
\end{center}
\end{figure}

In the simplest Paschen-Back linear Zeeman regime ($\beta << 1$), when the $m_l$ degeneracy is removed, all energy levels are shifted by an amount $\frac{1}{2}m_lh\omega_C$. A single line is split into a triplet composed of an unshifted central $\pi$ component ($\Delta m_l=0$), a red-shifted $\sigma_+$ ($\Delta m_l=+1$) component and a blue-shifted $\sigma_-$ ($\Delta m_l=-1$) component. The $\pi$ component occurs at the zero field frequency, $\omega_0$, while the two satellite $\sigma$ components occur at $\omega_0 + \omega_L $ and $\omega_0 - \omega_L$ where $ \omega_L=\omega_C/2$ is the Larmor frequency. The $\pi$ components are linearly polarised, while the $\sigma_-$ and the $\sigma_+$ components are circularly polarised with opposite signs when viewed along the magnetic field. In the linear regime and in wavelength units the splitting between a $\sigma$ component and the unshifted central $\pi$ component is
\begin{equation}
\delta \lambda_{L} =20.2 \left(\frac{\lambda}{6\,564\mbox{\AA}}\right)^2\,\left(\frac{B}{10^6\,{\rm G}}\right) ~\mbox{\AA}
\end{equation}
If the fields are low enough ($\sim 10^6 - 10^7$\,G) and the intrinsic broadening of the line due to pressure effects does not mask the Zeeman splitting, the characteristic pattern of a Zeeman triplet is detectable in flux spectra in the lower members of the Balmer and Lyman series at spectral resolutions typically used in WD surveys ($\approx 10$\,\AA). We show in Figure \ref{zeem_triplets_low} the spectra of seven SDSS DAP WDs \citep{Schmidt2003} showing the splitting of H$_{\alpha}$ into a classical Zeeman triplet.
At higher spectral resolutions, Zeeman triplet patterns can be detected in narrow non-LTE line cores in MWDs, extending the range of fields that can be measured by spectroscopy down to $\sim 10^5-10^6$~G. A nice set of data depicting how H$_\alpha$ varies as the magnetic fields strength increases is the series of low field WDs shown in Fig.\,\ref{Landstreet2016_Weak_Series} obtained by \citet{Landstreet2016}. 
A powerful method for measuring magnetic fields, particularly in the very low field regime where the Zeeman splitting is small and unresolvable, is the use of circular spectropolarimetry across lines. A statistically significant detection of the expected change in the sign of circular polarisation across spectral lines allows such fields to be measured, and this is the method used to discover kilo-Gauss fields (see Figure
\ref{Landstreet_LowField} from \citet{Landstreet2019} and also \S\,\ref{incidence}).

In the next field regime the quadratic term dominates in the Hamiltonian so that the $l$ degeneracy is also removed although the field is still not strong enough for adjacent $n$ manifolds to overlap significantly ($\beta < 10^{-3}$ for Balmer lines). In this regime the $\pi$ components are also shifted. The quadratic effect is expected to be first seen at low fields as an asymmetry in the line profile and a blue-ward displacement of the centroid of the line from its zero field position and then at higher fields with the individual components resolved. Note that the quadratic effect becomes stronger as $n_u$ (upper principal quantum number) increases for a given $n_l$ (lower principal quantum number) - that is, it is relatively more important in the higher members of a given series as shown in Fig.\,\ref{zeem_triplets_low}. The majority of MWDs are discovered through the detection of resolvable linear or quadratic Zeeman structures.
\begin{figure}[htbp]
\begin{center}
\includegraphics[width=10cm]{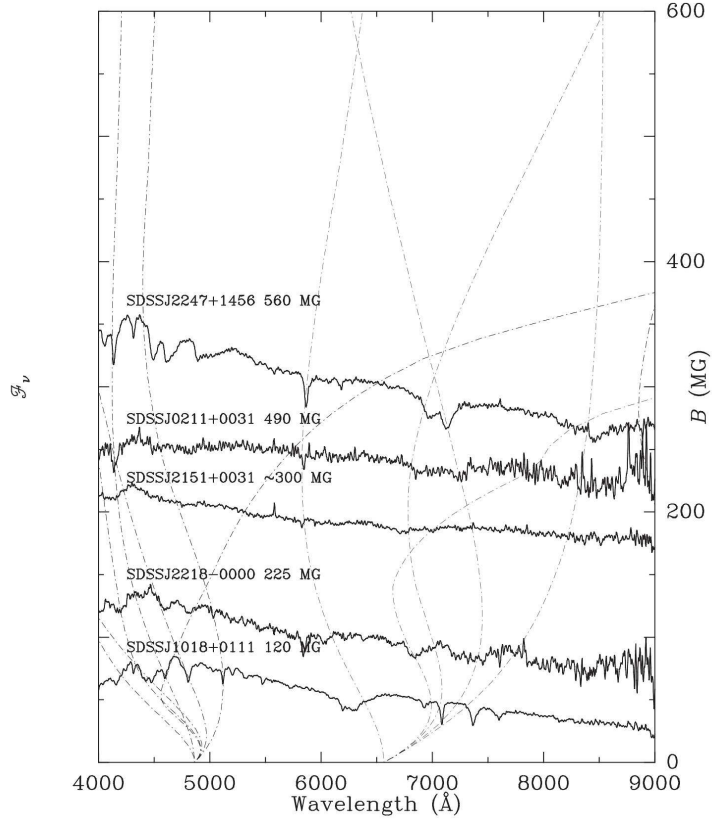}
\caption{The spectra of five MWDs with fields between 120 and 560\,MG showing stationary Zeeman components of H$_{\alpha}$ and H$_{\beta}$ \citep[from][]{Schmidt2003}. The dash-dotted lines are the Balmer lines transitions as a function of magnetic field strength in MegaGauss. All the strong absorption features observed in these spectra are associated with transitions whose wavelengths are stationary (or nearly stationary) with magnetic field strength.}
\label{zeem_triplets_high}
\end{center}
\end{figure}
At even higher fields ($\beta \sim 1$ - the diamagnetic regime), different $n$ manifolds begin to overlap and the only good quantum numbers are the parity $\pi$ and $m_l$. Although the energy level diagram shows no simple structure at these fields, in the strong field regime ($\beta >>1$ ) a new structure begins to appear, reflecting the quantisation of the motion of the electron perpendicular to the field into Landau energy states, with the motion along the field being effectively 1-D Coulombic. In the approach to this regime, the $\sigma^+$ components become nearly ``stationary'' in the sense that large changes in the magnetic field result only in small changes in the wavelength.  MWDs are discovered through the detection of these stationary components. In Figure \ref{zeem_triplets_high} we present a montage of spectra for a range of field strengths showing the appearance of stationary components \citep{Vanlandingham2005}.

\subsubsection{Forbidden lines}

\begin{figure}[htbp]
\begin{center}
\includegraphics[width=0.9\textwidth]{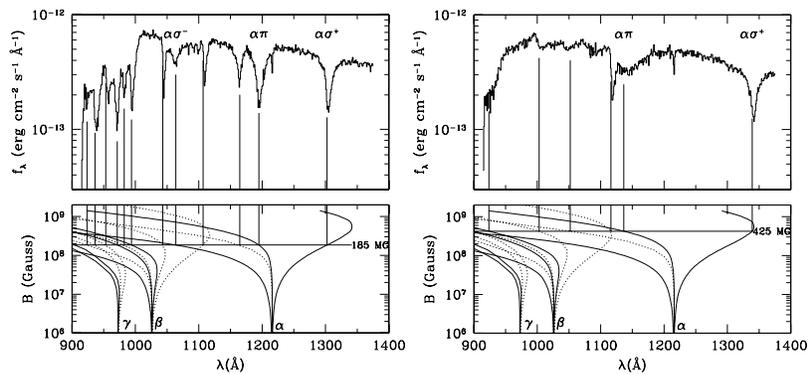} 
\caption{Top: Phase-resolved HST FOS/FUSE spectroscopy of RXJ\,0317-853 at polarimetric phase $\phi=0$
  (left) and $\phi=0.5$ (right). Bottom: the solid curves are the
  hydrogen Lyman transitions and the dotted curves the forbidden
  transitions that are enabled because of the strong electric fields
  induced by the super-strong magnetic fields. The observed spectra
  are compared to predicted line positions showing that the spectrum
  at $\phi=0$ has a field $B=185$\,MG, while at $\phi=0.5$ has a field $B=425$\,MG \citep[from][]{Vennes2003}.}
\label{lyman}
\end{center}
\end{figure}
The UV spectra of both isolated and accreting MWDs have shown the presence of transitions that are normally forbidden. For example, the $1s_0-2s_0$ Ly$_\alpha$ component that violates the parity/magnetic moment selection rules was detected by \citet{Burleigh1999} in the UV spectrum of the isolated HFMWD RXJ\,0317-853 and by \citet{Gaensicke01, Hoard2004} in the UV spectrum of the accreting MWD in the MCV AR\,UMa. Further observations of RXJ\,0317-853 (see Fig.\,5) conducted by \citet{Vennes2003} revealed the presence of many more forbidden transitions (seven in total) comparable in strength to permitted transitions. \citet{Burleigh1999} have explained that such transitions become possible in the dense atmospheres of MWDs because free electrons and ions can induce Lorentzian electric fields as strong as $10^9$\,V\,m$^{-1}$ that act perpendicularly to the magnetic field. Without these electric fields, the magnetic quantum number and the $z-$parity $\pi_z$ are conserved. However, only the magnetic quantum number is conserved for parallel electric and magnetic fields while only the $z-$parity is conserved for perpendicular electric and magnetic fields. If fields are randomly oriented there is no discrete symmetry and thus additional dipole transitions, such as $1s_0-2s_0$, are enabled \citep{Burleigh1999}.

\subsubsection {Helium lines}\label{HElines}

\begin{figure}[htbp]
\begin{center}
\includegraphics[width=10cm]{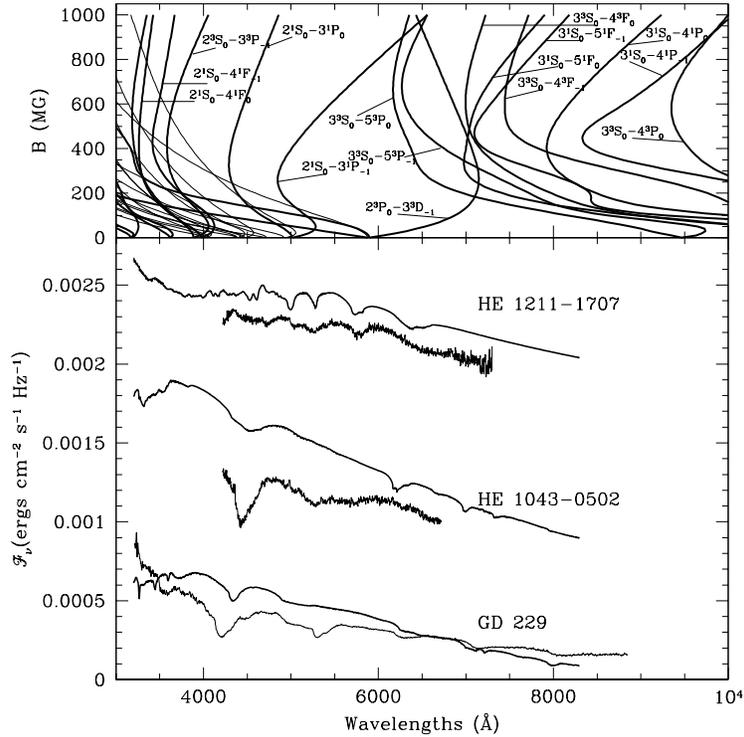}
\caption{Top panel: Helium transitions as a function of magnetic field strength in MegaGauss. Bottom panel: A comparison of centred dipole models for helium rich WDs with observations of GD\,229, HE\,1043-0502, and HE\,1211-1707. The models have, respectively, polar fields $B_d=$ 520, 820, and 50\,MG \citep[from][]{Wickramasinghe2002}.}
\label{helium_fits}
\end{center}
\end{figure}

MWDs in which helium is the dominant element in their atmospheres, the DBP (or DBH) MWDs, are characterised by Zeeman split HeI lines in their high temperature spectra. The understanding of their spectra required the unravelling of the structure of the energy levels of the two electron He atom at arbitrary fields. Following early numerical calculations of the wavelengths and transition probabilities of the stronger HeI lines in the low field regime ($\beta<<1$) \citep{Kemic1974}, detailed calculations carried out in the 1990's led to the determination of accurate energy levels of the low-lying states of HeI at arbitrary magnetic fields \citep{Becken2001, AlHujaj2003} and references therein). The calculations cover the lowest five ($n\le 5$) singlet and triplet states for the subspaces $m=0,\pm 1,\pm 2,\pm 3$ and include transition probabilities for the important optical and UV transitions. The lines split into triplets in the low-field regime, and stationary components appear as for hydrogen at high fields \citep[see][and Figure \ref{helium_fits}]{Jordan2001}.

Zeeman split HeI lines have been identified in a  sample of MWDs with fields in the range $4\times 10^6 - 6\times 10^8$\,G including the strongly polarised MWD GD\,229 \citep{Jordan1998, Wickramasinghe2002}. A comparison of the spectra of GD\,229, HE\,1211-1707, and HE\,1043-0502 with HeI Zeeman models constructed under various simplifying approximations is shown in Figure \ref{helium_fits} \citep{Wickramasinghe2002}. The identifications with HeI lines are secure except in the case of HE\,1043-0502 where other interpretations may be possible. The discrepancies between observations and theory in GD\,229 may be caused by resonances in the HeI bound-free opacities for which there is at present no adequate theory (see \S \ref{atmo}).

\subsubsection{Carbon molecular bands}\label{Carbon_lines}

A sub-class of cool WDs ($T_{\rm eff}<8,000$\,K) have atmospheres that (i) are He dominated, (ii) are H deficient, and (iii) exhibit molecular bands of C$_2$ that increase in strength as the temperature decreases. Occasionally they also show CH and atomic C lines. These WDs were first noted in the late 50's \citep{Greenstein1957_DQ} and were later named ``DQ'' WDs \citep{Greenstein1984_DQ}. They are too cool to allow the formation of neutral He lines and any trace of pollution reveals itself easily due to the high pressure and low continuum opacity of helium-dominated regions.  The presence of carbon in these cool DQ WDs is probably caused by the dredge-up from the core by the helium convection regions as first proposed by \citet{Pelletier1986}.

Spectropolarimetric observations have shown that a few of these cool DQ WDs are strongly magnetic \citep{Liebert1978_DQ}. More recent observations have revealed field strengths in the range $\sim 7-100$\,MG \citep[][and Kawka et al. 2019, in preparation]{Vornanen2010, Vornanen2013}.
\begin{figure}[htbp]
\begin{center}
\includegraphics[width=8cm]{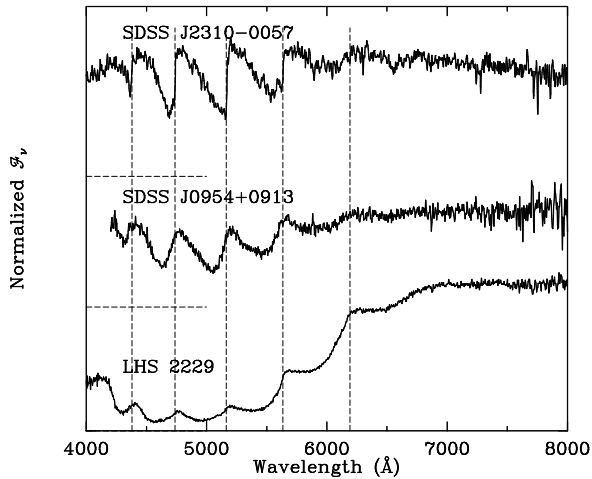}
\caption{Comparison between SDSS\,J0954+0913, the normal (C$_2$ Swan band) molecular WD SDSS\,J2310 0057, and the peculiar magnetic DQ star LHS\,2229. The horizontal dashed lines indicate the zero-flux levels while the vertical dashed lines indicate the band heads of normal C$_2$ Swan features \citep[from][]{Vanlandingham2005}.}
\label{swan_bands}
\end{center}
\end{figure}
We show in Fig.\,\ref{swan_bands} three WDs showing the distortion of the characteristic C$_2$ Swan bands with increasing magnetic field. The WD LHS\,2229 has a higher circular polarisation than SDSS\,J0954+0913 while SDSS\,J2310-0057 is unpolarised and non-magnetic. We note that \citet{Kowalski2010} showed that the blue shifted C$_2$ molecular bands can be explained by the high pressure that exists in very cool WDs. Therefore, these shifted Swan molecular bands are most likely C$_2$ molecular bands that have been pressure shifted.

Field estimates of DQ MWDs have so far relied on theoretical estimates of the quadratic Zeeman effect on C$_2$ and CH bands based on the simplified rigid rotator model of \citet{Bues1987}.  More detailed theoretical calculations of the Zeeman effect in the Paschen Back regime are available but have so far only been used to model the intensity and polarisation spectrum of G99-37 indicating a field of 8\,MG \citep{Berdyugina2007}. Models for the more strongly polarised and presumably higher field WDs, such as LHS\,2229, are still unavailable.

\subsubsection{Metal lines}\label{Metal_lines}

WDs have such a strong surface gravity that the diffusion timescale for metals to sink to the bottom of the atmospheres is very short \citep[days to weeks for a hydrogen-dominated atmosphere, as shown by][]{Koester2006}. Nonetheless, some $25$\% of WDs show traces of metals in their spectra (generally Ca and Si, but also Mg, Fe, Na and others, \citep{Zuckerman2003, Zuckerman2011, Farihi2011, Kawka2016, Farihi2016}. We shall come back to these intriguing objects in \S\,\ref{DZ}.
The level of an electron in an atom can be described by the three quantum numbers, $J$, $L$, $S$, where $J$ is the total angular momentum, $L$ is the orbital angular momentum and $S$ is the spin angular momentum. When a metal atom is placed in a magnetic field its levels are split into  $2J+1$ components that are defined by the magnetic quantum number $m=-J, -J+1,\cdots, J-1, J$. The resulting shifted Zeeman lines can be calculated using:
\begin{equation}
\Delta \lambda = \frac{eB\lambda^2}{4\pi m_e c^2}(g_l m_l - g_u m_u) \approx 4.67\times10^{-7} \lambda^2 B (g_l m_l - g_u m_u)
\end{equation}
The magnetic quantum numbers of the upper and lower levels are $m_u$ and $m_l$, respectively. Just like for the hydrogen atom, the permitted transitions that are allowed are $\Delta m = 0, \pm 1$. Similarly, the L\'ande factors of the upper and lower levels are $g_u$ and $g_l$, respectively. L\'ande factors for light elements, which includes calcium, sodium and magnesium, can be calculated assuming LS coupling:
\begin{equation}
g = 1+\frac{J(J+1)-L(L-1)+S(S+1)}{2J(J+1)},
\end{equation}
In heavier atoms, such as iron, the interaction between the orbital and spin angular momenta is as strong as the interaction between individual spins or orbital angular momenta and LS coupling is no longer valid. L\'ande factors for heavier elements can be obtained from the Vienna Atomic Line Database (VALD)\footnote{http://vald.astro.univie.ac.at/~vald3/php/vald.php \citep{Kupka2000}.}

\begin{figure}[htbp]
\begin{center}
\includegraphics[width=0.7\textwidth]{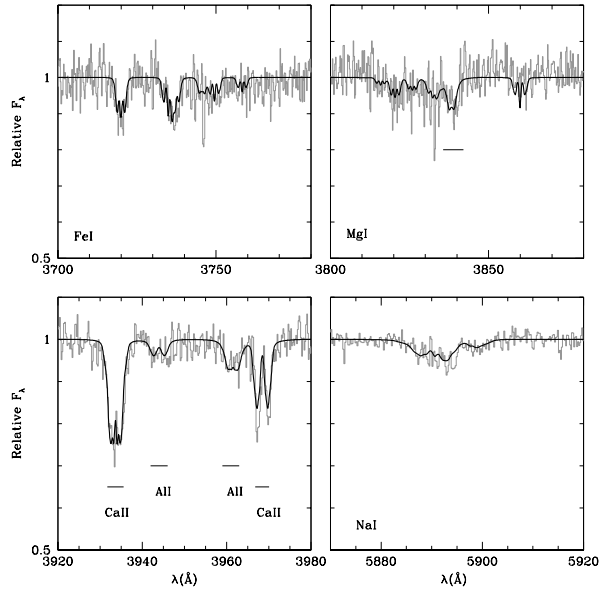}
\caption{Spectrum of the cool DAZ NLTT\,7547 (grey) compared to the best fitting model (black) at $T_{\rm eff} = 5460$\,K, $\log{g} = 8.04$ and $B_d=240$\,kG. The Zeeman split lines of sodium, magnesium, calcium and iron are shown. \citep[see][]{Kawka2019_DZ} for details.}
\label{fig_nltt7547}
\end{center}
\end{figure}

The relative intensities of the Zeeman components in metal lines is proportional to functions of the total angular momentum and the magnetic quantum number \citep{Condon1963}. For transitions where $\Delta J=0$ and $\Delta m=0$
\begin{equation}
    I \propto m^2,
\end{equation}
and for $\Delta m=\pm1$
\begin{equation}
I \propto \frac{1}{4}(J\mp m)(J\mp m+1).
\end{equation}
For transitions where $J \rightarrow J+1$ and $\Delta m=0$
\begin{equation}
I \propto (J+1)^2-m^2,
\end{equation}
and for $\Delta m=\pm1$
\begin{equation}
I \propto \frac{1}{4}(J\pm m+1)(J\pm m+2).
\end{equation}
Finally, for transitions where $J \rightarrow J-1$ and $\Delta m=0$
\begin{equation}
I \propto J^2-m^2,
\end{equation}
and for $\Delta m=\pm1$
\begin{equation}
I \propto \frac{1}{4}(J\mp m)(J\mp m-1).
\end{equation}
Table\,\ref{tbl-split1} in Appendix A lists three different angular momenta, L\'ande factors and relative intensities of the Zeeman components for the most commonly detected element lines in WD atmospheres. Fig.\,\ref{fig_nltt7547} shows the spectrum of NLTT\,7547, a cool DAZ WD, compared to model spectra calculated at $B_d=240$\,kG. The different strengths of the multiple Zeeman components of the metal lines are clearly noticeable.

The regime described above is valid for heavy elements in relatively low magnetic fields. In higher magnetic fields, the Paschen-Back regime becomes appropriate and the Zeeman split lines become triplets. For Na, this occurs for $B\gsimeq 100$\,kG\, and for others such as Mg and Ca, this occurs for $B\gsimeq 2$\,MG.

\subsection{Cyclotron lines and magnetised winds}\label{cyclotron}

Free electrons in the atmospheres of MWDs are expected to give rise to thermal cyclotron radiation which results in an absorption feature at the cyclotron fundamental $\omega_{C}$ which corresponds to a wavelength
\[
\lambda_c=\frac{2\pi}{\omega_C}=10\,710 \left( \frac{10^8\,{\rm G}}{B}\right) ~ 
\mbox{\AA}
\]
Although the integrated cross-section is similar to that for bound-free H opacity, the line is intrinsically narrow with a thermal width 
\[
\delta \omega_C/\omega_C =\sqrt{2kT/m_ec^2}\vert\cos \psi\vert
\]
where $\psi$ is the angle between the field direction and the line of sight \citep{Martin1979a}.  The cyclotron fundamental should in principle be detectable in the optical to near IR band in MWDs with fields of a few $10^8$\,G.

So far there has been no clear evidence for such a feature in the spectra of high-field MWDs, although the broad absorption feature extending over $\sim 1,000$\,\AA\ in the optical band in the strongly polarised MWD G240-72 may be of cyclotron origin \citep{Martin1979a}. In general, the detection of cyclotron features requires a very nearly uniform field structure over the visible surface because magnetic field broadening (with $\lambda \propto B$) is very effective in rendering the feature undetectable \citep[see][]{Martin1979a}.  Given the complex field structures that are now found in MWDs, and the possibility that the visible surface may present a nearly uniform field at some magnetic phases, one may expect that the cyclotron feature will be unambiguously detected in the absorption spectra of some MWDs in the near future.

Another method of detecting the cyclotron resonance is by using polarimetry. Cyclotron emission is 100\% circularly polarised when viewed along the field and elliptically polarised at a general viewing angle. The polarisability of the plasma and the index of refraction changes across the cyclotron resonance so as to give rise to a rotation in the polarisation angle by $90^\circ$ \citep[or less than $90^\circ$ if vacuum polarisation is important, see][]{Zheleznyakov1991}. Spectropolarimetric data of Grw+70$^\circ$8247 do show such a rotation near 5500\,\AA \citep{West1989}, although such rotations may also be caused by continuum absorption edges.

The intriguing possibility that radiation pressure acting on the cyclotron resonance may drive outflows in high-field MWDs has been discussed by \citet{Zheleznyakov1991}. The mass loss rates are estimated to be $\sim (0.3-1)\times 10^{-13}$\,M$_\odot$\,yr$^{-1}$ for MWDs with effective temperatures $(2-5)\times 10^4$\,K and magnetic fields of $(1.5-10)\times 10^8$\,G. At the low densities expected in such coronal outflows, vacuum polarisation could have a significant effect across the vacuum resonance where the polarisability of the plasma and the vacuum cancel \citep{Gnedin2006, Zheleznyakov1991}. The vacuum resonance occurs at $\lambda \propto B/N_e$ and in the context of WDs, will be important in the optical band at the low electron densities $N_e \le 10^8$cm$^{-3}$ that may be expected in such outflows.

\subsection{Constraining gravity theories using gravitational birefringence} \label{gravity}

Some non-standard theories of gravity that go beyond general relativity by coupling the electromagnetic field directly with gravitational gauge fields could be constrained by studies of continuum polarised emission from MWDs \citep{Solanki1999}. For certain types of coupling, space-time becomes birefringent (gravitational birefringence) causing a phase-shift between different polarisation components leading to an additional source of depolarisation over and above the depolarisation that occurs due to the magnetic bi-refringence of the plasma as the radiation propagates through the atmosphere. The predicted phase-shift is  directly proportional to the coupling constant and the gravity at the surface of the star and inversely proportional to the wavelength of observation. A measurement of the excess depolarisation of polarised radiation emitted by an astronomical source can in principle be used to constrain such theories.  The observed levels of polarisation in the MWDs Grw+70$^\circ$8247 \citep{Solanki1999} and RE\,J0317-853 \citep{Preuss2004} have been used to place some limits on the coupling constant for specific non-metric gravity models.  These limits could be significantly improved when we have a better understanding of the atmospheres of MWD atmospheres and their intrinsic polarised emission. Potentially, neutron stars with their stronger surface gravities would be better suited for constraining such theories.

\section{Model atmospheres and magnetic field structure}\label{atmo}

\subsection{Model atmospheres}

Calculations of model atmospheres for MWDs must include the effects of the magnetic field on continuum and line opacities and incorporate a full solution of the polarised transfer equations in the four Stokes parameters $\{I,Q,U,V\}$ that allows for the mixing of the different modes of propagation due to magneto-optical effects \citep{Martin1979a}.  Magnetic fields may also have a direct influence on the hydrostatic equilibrium of the atmosphere. However, there are no calculations at present that fully account for all of these effects so our understanding of the atmospheres of MWDs is still incomplete.

\subsubsection{The force-free approximation and hydrostatic equilibrium} \label{forcefree}

The possible effect that a magnetic field may have on the hydrostatic and thermal equilibrium of a star has been discussed by several investigators  \citep[e.g.][]{Landstreet1987, Fendt2000}.  It is generally assumed that the magnetic field in the atmosphere is force-free maintained by currents in the interior of the WD. In this approximation, the magnetic pressure (${B^2}/(8\pi)$) balances the magnetic tension along field lines ($\vec{B}\cdot\nabla{\vec B}$) so that there is no net magnetic force, even though each of these forces exceeds the gas pressure in the atmosphere by many orders of magnitude.

It has been argued that the Ohmic decay of a fossil field would lead naturally to a decay-induced Lorentz force that would drive a meridional motion perpendicular to field lines and affect the magneto-hydrodynamic structure of the WD atmosphere \citep{Landstreet1987, Jordan1992}.  Non force-free configurations are also possible if the electric currents that maintain the field in the interior of the star also extend to the atmosphere \citep{Fendt2000}.  It has been estimated that a $\sim 2-10$\,MG toroidal field that is superimposed on a 100\,MG poloidal field and maintained by current flows would lead to an increase in the scale height of the atmosphere by a factor $\sim 10$. Such a change in the atmospheric structure should in principle be detectable from line profiles \citep[see][]{Friedrich1994}.

\subsubsection{Magnetic dichroism and birefringence} \label{dichroism}

Field dependent opacities (magnetic dichroism) and the associated magneto-optical parameters (magnetic birefringence) can be calculated for bound-bound transitions of hydrogen or neutral helium using existing calculations of energy levels and transition probabilities. As we have seen, the bound states of H and He show a rich and complex structure which must translate into complex structure in the bound-free opacities. A full quantum mechanical treatment of bound-free transitions must allow for the quantisation of the continuum into quasi-Landau levels in the presence of the Coulomb potential. The first results of such calculations for the hydrogen atom show how complex resonances develop in the Landau continua \citep{Merani1995}.  Attempts at using these results to model the flux and polarisation spectra of Grw+70$^\circ$8247, however, have met with limited success particularly with regard to polarisation. When the H bound-free opacities calculated by \citet{Zhao2006} and covering a wider field range are supplemented with associated magneto-optical parameters they should lead to the construction of better models of the hotter magnetic DA WDs. At present, though, there are no similar calculations for the helium atom, which are important for the modelling of the magnetic DB WDs. Nor are there calculations of the field dependence of $\rm{H}^-$ or $\rm{He}^-$ opacity which are important in cool WDs.

Although bound-free calculations predict narrow resonances in the Landau continua, there has been no observational evidence for such features in the spectra of MWDs. As with the cyclotron resonance, because of magnetic field broadening, the detection of such features would require a field structure which is very nearly uniform over the visible stellar surface.

The model atmospheres that are currently in use for studying MWDs have adopted various approximations to the continuum opacity.  These range from a simple linear splitting and shifting of ionisation edges, to a splitting and shifting of edges which allow transitions from generally Zeeman-split bound states to discrete continuum states assumed to be the Landau's states of a free electron \citep{Jordan1992} for the case of hydrogen \citep[see][for more details]{Wickramasinghe2000}. Such approximations have proven to be adequate for establishing or constraining the magnetic field structure over the stellar surface where the dominant effect is the Zeeman splitting and shifting of the lines as a function of field strength (\S \ref{fieldstruc}).

\subsection{Magnetic field broadening and magnetic field structure} \label{fieldstruc}

The wavelength of a Zeeman component is strongly field dependent in most Zeeman regimes except for those components that become ``stationary'' in the high-field regime (see section \ref{measure} and Fig.\,\ref{zeem_triplets_high}). In addition to pressure broadening, the spectral lines are also broadened by field spread across the visible stellar surface. In the simplest case of a dipolar field distribution and a component which shifts entirely according to the linear Zeeman effect, magnetic field broadening will impart to the line an intrinsic width of $\approx 2 \lambda_{L}$, where $\lambda_{L}$ is the Larmor wavelength (see \S\,\ref{Hlines}). For typical field strengths of $\approx 10^7$\,G, and WD temperatures of $\approx 10,000$\,K, this width exceeds the pressure or thermal width of a line ($\approx 10$\,\AA) by an order of magnitude, and thus magnetic field broadening dominates. Stationary components can therefore be used to investigate field structure even in the absence of a detailed atmospheric structure.

Rotating WDs provide the best opportunity for investigating the underlying field structure since they allow different parts of the stellar surface to be viewed at different rotational (and generally magnetic) phases. The Zeeman shift of a line component depends only on the magnitude of the magnetic field, while the circular and linear polarisation of a Zeeman split component also carry information on the field direction.  Phase-dependent spectropolarimetric observations therefore provide strong constraints on field structure. With the advent of large optical telescopes, it has recently become possible to carry out such studies for a few stars, although at present the data have been restricted to circular spectropolarimetry which is a major limitation.

All studies assume that the magnetic field in the atmosphere is force-free and derivable from solutions of Laplace's equation (see however \S\,\ref{forcefree}).  Two approaches have been adopted. The first assumes a truncated multipolar expansion using spherical harmonics that include only the lowest harmonics (usually $l=2$ (dipole), $l=3$ (quadrupole), $l=4$ (octupole) with $m=0,..l$).  The series is arbitrarily truncated for computational efficacy which leads to restrictions on the type of field structures that can be modelled. The second approach, for which there is also no obvious physical justification, assumes that the field structure can be described by a combination of off-centred dipole, quadrupole and octupole zonal ($m=0$) components, with arbitrary tilt angles and offsets for each of the components \citep[see][]{Euchner2002}.

\begin{figure}[htbp]
\begin{center}
\includegraphics[width=0.9\textwidth]{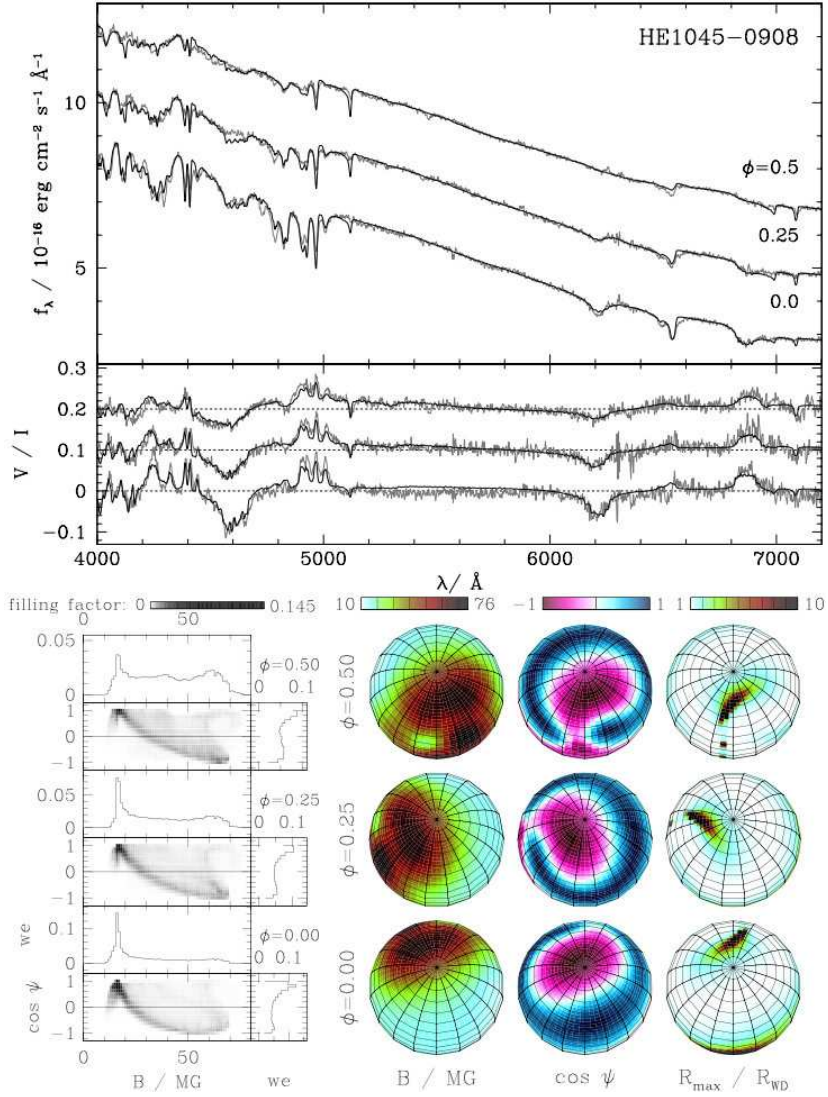}
\caption{Zeeman tomographic analysis of the field structure of HE\,1045+0908 \citep[from][]{Euchner2005} using an off-centred, non-aligned combination of dipole, quadrupole, and octupole. Top: Flux and polarisation data overlapped with the best fit model. Bottom left: $B-\psi$ diagram, where $\psi$ is the angle between the local field direction and the line of sight. Bottom right: absolute value of the surface magnetic field, cosine of $\psi$, and the maximum radial distance reached by field lines in units of the WD radius \citep[see][for details]{Euchner2005}.}
\label{he1045}
\end{center}
\end{figure}

Detailed studies have been carried out for HE\,1045+0908 and PG\,1015+014 \citep{Euchner2005, Euchner2006}. The best fit to the field structure of HE\,1045+0908 is shown in Fig.\,\ref{he1045} \citep{Euchner2005}.  For this star, the Zeeman tomographic analysis has revealed a field structure which is dominated by a quadrupole, but which also contains additional dipole and octupole contributions. The field structure in PG\,1015+014 appears even more complex. Model fits using multipolar expansions with $l=1,2,3,4$ including tesseral components ($m\neq 0$) give reasonable agreement with observations but also indicate that more terms are needed. The quality of the model fits appears to depend generally on which class of model is adopted, highlighting the fact that the inverse problem is not as yet adequately constrained. Stronger constraints will require improvements in dealing with the inverse problem, and the use of all four Stokes parameters.

\subsection{Convective mixing and accretion}\label{mixing}

The hydrogen that is seen in the atmospheres of non-magnetic WDs could either be accreted, or be remnant from the time of formation. Sub-photospheric convection could mix in the outer H rich layers in certain ranges of effective temperature, provided they are sufficiently thin, and result in a transformation of the spectral appearance from DA to DB as first suggested by \citet{Strittmatter1971}. The surface atmospheric composition could also be altered by interstellar accretion or circumstellar accretion from a debris disc, and modified by gravitational diffusion, and the current evidence is that all these processes (accretion-diffusion-convective mixing) are in operation \citep[see, e.g.,][]{VauclairGreenstein1979,WesemaelTruran1982}.

Magnetic fields can affect each of these processes to different extents. They may inhibit interstellar accretion or modify the accretion pattern by shielding the star from incoming plasma. The ratio $\zeta$ of ram pressure to magnetic pressure for a non-rotating star with a dipolar field moving with a velocity $v$ through an ambient medium of density $n_{\infty}$ can be estimated to be \citep[e.g.,][]{Strittmatter1971}:
\begin{equation}
\zeta \sim \left(\frac{n_{\infty}}{{\rm cm}^{-3}}\right)\left(\frac{M}{M_\odot}\right)^6\left(\frac{4\times 10^7 {\rm G}}{B}\right)^2\left(\frac{10^9\,{\rm cm}}{R}\right)^6\left(\frac {50\,{\rm
km\,s}^{-1}}{v}\right)^2.
\end{equation}
In the strong field case ($\zeta \le 1$) we expect an inhibition of accretion, while in the weak field case ($\zeta \ge 1$) we may expect preferential accretion of ionised material on to the polar regions of the star. Given the strong dependence of $\zeta$ on $B$ we may expect patchy H/He surface compositions in an otherwise He rich atmosphere due to differential effects on accretion in stars with fields of $\sim 1-4\times 10^7$\,G.

It is possible that a MWD with a field that is not strong enough to inhibit accretion ($\zeta \ge 1$) if it is non-rotating, may inhibit accretion by the propeller mechanism if it does rotate.  This mechanism could be important even in MWDs with modest magnetic fields ($\sim 5\times 10^7$\,G) and rotation velocities ($\gsimeq 20\,\rm{km}\,\rm{s}^{-1}$).

A magnetic field may have a more direct effect on the atmospheric composition by inhibiting mixing of the outer layers of a WD with a convective envelope. This process may lead to the development of patchy surface chemical compositions even in weaker field stars. The work of \citet{Valyavin2014} indicates that magnetic fields may quell atmospheric convection and cause the formation of dark spots. At the high end of the field distribution convection can be suppressed altogether over the entire stellar surface thus altering the cooling evolution by making MWDs appear younger than they are. Using 3D radiation magneto-hydrodynamic simulations, \citet{Tremblay2015} confirmed that convection is suppressed by magnetic fields, however they also show that the cooling of the WD is not affected until the convective zone couples with the degenerate core, which occurs at effective temperatures below 6000\,K. \citet{Tremblay2018} provided evidence that magnetic fields suppress convection by modelling the ultraviolet and optical spectra of the MWD WD\,2105$-$820, which has an effective temperature of 10\,000\,K and should have a convective atmosphere. In their analysis, they showed that the ultraviolet and optical spectra resulted in consistent atmospheric parameters if the spectra were fitted with non-convective models. On the other hand, ultraviolet and optical spectra of non-magnetic WDs of similar temperature, had consistent atmospheric parameters when convective models were used. Given the above considerations, one may expect differences in the atmospheric properties of non-magnetic and MWDs, and there is some evidence that this may indeed be the case.

Feige\,7 is rare among MWDs in showing both H and He lines in its spectrum. The spectrum varies with a rotation period of 2.19\,hrs and can be interpreted as evidence for a patchy surface chemical composition with some regions being more hydrogen rich than others \citep{Achilleos1992b}.  The star has an effective temperature of $\sim 21,000$\,K in the regime where convective mixing is expected to take place.

In addition to spectral line variations, rotating MWDs are expected to also exhibit variations in continuum flux and polarisation as the underlying magnetic field is viewed at different magnetic phases. Such a behaviour is typified again by Feige\,7 \citep[see][]{Wickramasinghe2000} that exhibits variations at a level of 4\% in V which can be explained in terms of a patchy surface composition and the field dependence of continuum opacities (magnetic dichroism).

Flux variations at the 0.2\% level has been detected in the low-field MWD WD\,1953-011. This star has an effective temperature of $\sim 7900$\,K and a convective atmosphere \citep{Maxted2000, Brinkworth2005}. The spectral line variations have been interpreted in terms of a two component model consisting of an underlying dipolar field distribution with $B_d= 7\times 10^4$\,G and a spot with a field of $5\times 10^5$\,G covering $\sim 10$\% of the stellar surface \citep{Maxted2000}. The magnetic field strength appears too weak to explain the photometric changes purely in terms of magnetic dichroism \citep[see][]{Wickramasinghe2000}. Larger scale changes in opacity due to variations in the atmospheric He/H ratio across the surface may be a more plausible explanation, although unlike with Feige\,7, the star is too cool to show helium lines. Another possibility is that chemical inhomogeneities play a secondary role and that the magnetic field has a direct influence on the thermal structure of the atmosphere by inhibiting convective motions perpendicular to field lines. The changes in atmospheric structure may then lead to small scale variations in the optical flux over the stellar surface \citep{Brinkworth2005}.

\subsection{Planets around magnetic WDs}\label{planets}

In this section we shall focus specifically on the detection of planets using methods that are contingent on the presence of a magnetic field in the WD. 

GD\,356 is so-far the only WD to show the Balmer series in \emph{pure emission}.  The emission lines are split into Zeeman triplets and the lack of significant field broadening suggests that the lines originate from a region of the stellar surface with a nearly uniform field of $13\pm 2$\,MG  \citep{Ferrario1997_GD356}. The appearance of emission rather than absorption lines can be explained if there is a temperature inversion in the outer layers that begins at significant optical depth.  Nearly sinusoidal low amplitude (0.2\%) photometric V band variability has been detected in this star, yielding a rotation period of $\sim 115$\,min \citep{Brinkworth2004}.

The origin of the chromospheric type temperature inversion in GD\,356 has received much attention. One possibility is that the atmosphere is heated by X-ray emission from a corona, but searches for such emission have led to negative results with an upper limit of $6\times10^{25}$\,erg\,s$^{-1}$ to the X-ray luminosity being placed from Chandra observations \citep{Weisskopf2007}, an order of magnitude lower than what is required to explain the luminosity in Balmer line emission \citep{Ferrario1997_GD356}. The absence of detectable X-ray emission also argues against heating caused by accretion from a brown dwarf companion or the interstellar medium. Further studies by \citet{Wickramasinghe2010} using Spitzer Infrared Array Camera observations have confirmed this by placing a strong upper limit of 12\,M$_{\rm J}$ (where M$_{\rm J}$ is the mass of Jupiter) on the mass of a possible companion. In view of this result, models invoking accretion from a companion star can be dismissed.

\begin{figure}[htbp]
\begin{center}
\includegraphics[width=0.7\textwidth]{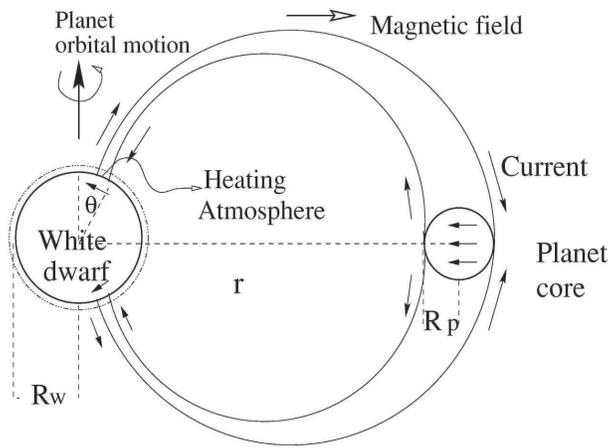}
\caption{The highly conductive iron core of a planet acts like a battery by generating electric currents as it moves in the magnetosphere of the MWD. The electric circuit that is setup causes the heating of the MWD’s atmosphere (the circuit's resistor) near its magnetic poles causing a temperature inversion and the formation of emission lines \citep[see][for details]{Li1998a}.}
\label{GD356planetLi1998}
\end{center}
\end{figure}

Another intriguing possibility is that GD\,356 has an Earth-type planetary companion that causes the heating of the atmosphere of the WD by the unipolar inductor mechanism \citep[see][who first suggested this possibility]{Li1998a}. On this model, the resistive dissipation of electrical currents that flow between the conductive iron core of an Earth-like planetary companion (stripped of its mantle during the red giant evolution of its star) and the MWD results in the heating of the upper atmosphere of the MWD. The model is similar to that proposed for the Io-Jupiter system and predicts the existence of heated regions near the footpoints of closed field lines which is generally consistent with the constraints provided by the Zeeman modelling. A schematic diagram illustrating this model is shown in Fig.\,\ref{GD356planetLi1998} \citep{Li1998a}.

The fate of planetary systems around main sequence stars that evolve into WDs is not well understood. Rocky planets that are dragged in during the red-giant phase of evolution of the parent star may be either fully encompassed by the star or their cores may survive in close orbit around the WD. According to \citet{Willes2005}, $\sim 0.1$\% of main-sequence planetary systems may evolve into MWD - Earth type planet systems. These planets may be detectable via the unipolar inductor mechanism before they are tidally disrupted.  A subset of these systems may emit radio emission by the electron-cyclotron maser mechanism at levels that may be detectable in future radio surveys in the $\sim 50-500$\,GHz band \citep{Willes2005}.

It may be possible to detect brown dwarfs and gaseous giant planets around MWDs through the detection of cyclotron emission from the MWD. For instance, the system SDSS\,J1212+0136 \citep{Schmidt2005} has an orbital period of 88.4\,min and consists of a L7 brown dwarf \citep{Debes2006} orbiting a MWD with a field of $\sim 7$\,MG. The cyclotron emission from the accretion shock on the surface of the MWD is powered by \emph{wind} accretion (as opposed to Roche lobe overflow) from its brown-dwarf companion \citep{Schmidt2005}. The strength of these cyclotron lines, and hence their detectability, depends on the specific accretion rate, but it is not clear whether Jupiter-like giant planets have winds of sufficient strength to produce measurable cyclotron emission.

There have been reports of the detection of planets in a number of CVs, among which the short-period DP\,Leo (a MCV that belongs to the polar sub-class) by \citet{Beuermann2011} who found convincing evidence that the third body orbiting this system is a giant planet in an elliptical orbit. This suggests that the occurrence of planets or planetary systems in post-CE binaries may not be a rare incidence. However, it is unclear whether these planets are first generation, that is, they were born together with their star in a proto-planetary disc, or second generation, that is, born from the material that was ejected during the AGB super-wind phase and/or common envelope evolution. Some 25\% of WDs show the presence of metals in their spectra caused by ongoing accretion of planetary debris (see \S\,\ref{DZ}). This and the detection of gaseous debris discs around many WDs strengthen the hypothesis that planets exist or did exist around a large fraction of WDs.

\section{Population properties of magnetic WDs}\label{pops}

\subsection{The incidence of magnetism}\label{incidence}

MWDs with fields in the range $10^6 - 10^9$\,G are readily recognised even at low spectral resolution ($\sim 10$\,\AA) and thus many MWDs have been discovered in surveys such as the Sloan Digital Sky Survey (SDSS), the Hamburg/ESO survey and the Edinburgh-Cape survey. Interestingly, the early work of \citet{Putney1997} revealed that the incidence of magnetism among WDs with no detectable absorption features (the DC WDs) is close to 22\%, which is much higher than among the general population of WDs. Further studies conducted over the following years revealed that a significant fraction of the magnetic DC discovered in this survey were in fact cool DAP white dwarfs. A similar survey conducted on the current much larger sample of DC WDs may lead to the discovery of a considerable number of new MWDs.

Higher resolution spectroscopic studies capable of measuring even lower fields have usually been limited to a few bright WDs. The high resolution ($\sim 0.1$\AA) ESO Supernova Ia Progenitor Survey (SPY) of more than 1,000 WDs presented an opportunity for the study of magnetism in the range $\sim 10^5-10^6$\,G. However, \citet{Koester2009}) reported that SPY only detected nine MWDs with fields below $10^6$\,G (five new and four already described in the literature), a result that seems to be generally consistent with spectropolarimetric surveys also sensitive to this field range (see below). 

Despite the large increase in the number of newly discovered MWDs, the question of the incidence of magnetism among WDs remains unresolved. Early studies based on magnitude limited surveys, such as the PG survey, indicated that $\sim 2$\% of WDs are magnetic. However, such surveys were biased against the discovery of massive and hence fainter WDs. A re-analysis of the data of the PG survey taking into account the different volumes that are sampled by different mass WDs resulted in an upward revision of the percentage of MWDs to at least $\sim 10$\% \citep{Liebert2003}. This revision rests on the observation that  MWDs with fields $\ge 10^6$\,G tend on average to have higher masses than their non-magnetic counterparts (see \S\,\ref{masses}). 

A complementary approach has been to isolate and study the properties of nearly complete volume limited samples of nearby WDs using field determinations from spectroscopy and spectropolarimetry. For the $13$\,pc sample, it has been estimated that $21\pm 8$\% of all WDs have magnetic fields greater than $\sim 3$\,kG, while for the $20$\,pc sample, an incidence of $13 \pm 4$\% has been estimated \citep{Kawka2007}. In this statistical analysis, however, it is suggested that the $20$\,pc sample is incomplete at the kilo-Gauss level. The more recent work of \citet{Holberg2016} corroborated these earlier findings by showing that the inner 20\,pc volume contains 12\% MWDs with a space density of $\sim 0.6\times10^{-3}$\,pc$^{-3}$.

The search for low-field MWDs ($B<10^6$\,G) is based almost exclusively on spectropolarimetric surveys, except for the few bright WDs for which high resolution spectra are available. A survey of 169 mainly northern WDs \citep[][and references therein]{Schmidt1995} led to the discovery of the first two low field WDs: LP\,907-037, with a longitudinal field $B_l = 8.5\times 10^4$\,G, and G\,217-037, with $-100\pm 15\le B_l \le +9\pm 12$\,kG. An extension of this survey to include 60 southern hemisphere WDs \citep{Schmidt2001b} led to the discovery of another possible low-field MWD with a longitudinal field of $6.1\pm 2.2$\,kG. Other spectropolarimetric studies with sensitivities capable of detecting kilo-Gauss fields resulted in the discovery of many more low-field MWDs. However, the sample of low-field MWDs, based on detections of polarisation across spectral lines, is still not large enough to establish the real incidence of very weak fields in WDs \citep{Landstreet2012, Bagnulo2018}. 

\begin{figure}[htbp]
\begin{center}
\includegraphics[width=7cm]{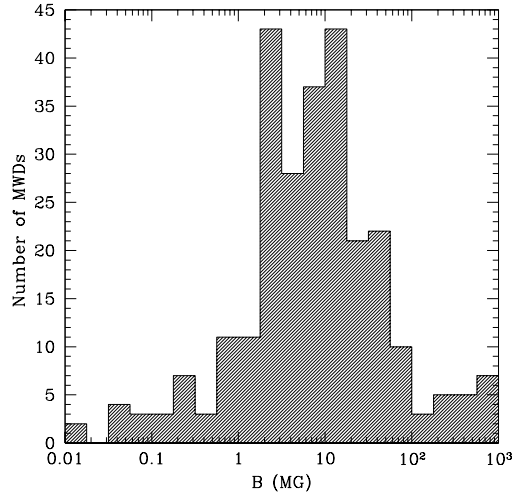}
\caption{The magnetic field distribution of all isolated MWDs (this work).}
\label{field_distribution}
\end{center}
\end{figure}

The current evidence is that the magnetic field distribution of WDs has a well defined high field component ($10^6-10^9$\,G, forming the class of high-field MWDs) whose incidence is about 3\% in magnitude-limited surveys, and a low field component ($<10^6$\,G, forming the class of low-field MWDs) whose incidence and properties are not as well established at the present time. The magnetic field distribution of all known MWDs is shown in Figure \ref{field_distribution}. 

\subsection{The Mass distribution}\label{masses}

The problem concerning the determination of surface gravities and masses from the line spectra of MWDs has always been intractable except in low field stars, i.e., $B \le \mbox{a~few} \times 10^6$\,G.  In this regime it is assumed that the magnetic field does not affect the atmospheric structure so that field broadening is small and existing Stark broadening theories (at zero magnetic field) can be used to model the line wings \citep[e.g.][]{Ferrario1998a}.

In principle, it should also be possible to use stationary components that are insensitive to field structure (see \S\,\ref{fieldstruc}) to estimate gravities from line profiles. However, this strategy must await a full theory of Stark broadening in the presence of crossed electric and magnetic fields, and the construction of more realistic model atmospheric structures that allow for the effects of the magnetic field. Although this is an important and unsolved problem pertaining to the physics of MWDs, their masses can now be found thanks to the parallax measurements provided by the Gaia mission. It is encouraging to see that the mass determinations for low field MWDs that were made through spectroscopic studies compare well with those provided by the parallax measurements of GAIA. Furthermore, the masses of MWDs that are determined through parallax will help us constrain Stark broadening theories (and atmospheric models) that allow for the presence of strong fields.

\begin{figure}[htbp]
\begin{center}
\includegraphics[width=10cm]{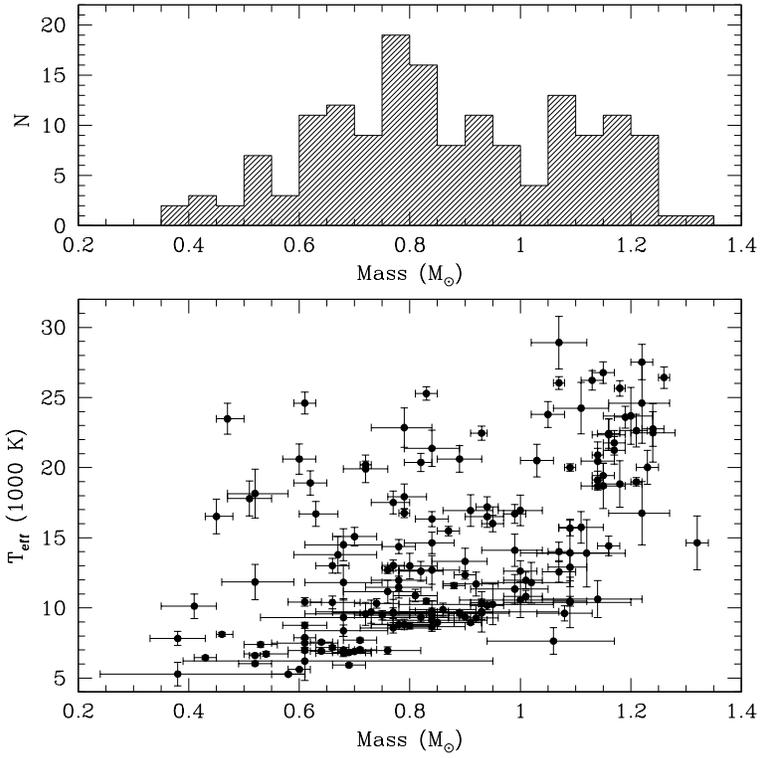}
\caption{Top panel: histogram of the mass distribution of hydrogen-rich MWDs that have SDSS photometry and Gaia parallaxes. Bottom panel: Effective temperature of these MWDs against masses (this work).} 
\label{Gaia_masses}
\end{center}
\end{figure}

\begin{figure}[htbp]
\begin{center}
\includegraphics[width=10cm]{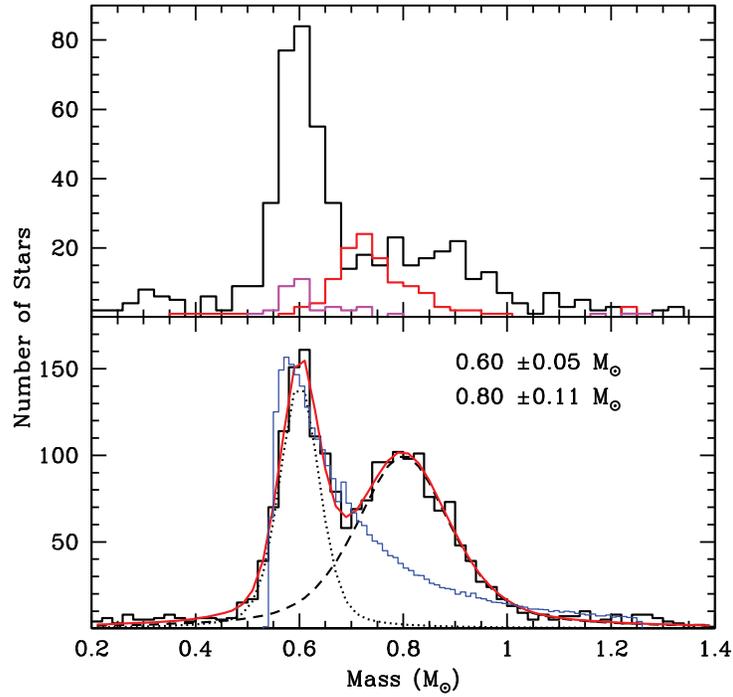}
\caption{Top panel: mass distributions of SDSS DA, DB and DC non-magnetic WDs, depicted in black, magenta, and red respectively \citep[see][for full details]{Kilic2018}. Bottom panel: mass distribution of the 100\,pc non-magnetic WDs with $T_{\rm eff}\ge 6,000$\,K. The presence of two distinct populations peaking near 0.6\,M$_\odot$ and 0.8\,M$_\odot$ is clearly evident \citep[see][for full details]{Kilic2018}.} 
\label{Kilic2018_masses}
\end{center}
\end{figure}

We show in Fig.\,\ref{Gaia_masses} the mass distribution of hydrogen-rich MWDs that have SDSS (DR12) photometry and Gaia trigonometric parallax. 
The data in Figure \ref{Gaia_masses} confirm that the MWDs as a group have a higher than average mean mass as first noted by \citet{Liebert1988}. The discovery by \citet{Kilic2018} that non-magnetic WDs show two very distinct populations, one peaking near  0.6\,M$_\odot$ and the other near 0.8\,M$_\odot$, is very surprising (see Fig.\,\ref{Kilic2018_masses}). The explanation of \citet{Kilic2018} is that the more massive population is the likely result of stellar merging events. However, a word of caution is in order, since \citet{Bergeron2019} noted that some of these high mass WDs could be $\sim 0.6$\,M$_\odot$ WDs with a mixed H/He atmosphere. Thus, the magnitude of this second high mass peak can only be established by conducting further observational studies to determine the atmospheric composition of this group of stars. Nonetheless, the existence of a large population of massive merged WDs is consistent with the studies of \citet{Holberg2016} of the WDs in the 13\,pc volume-limited sample. They find that the incidence of WDs in systems containing two or more stars is only 26\%. As first noted by \citet{Ferrario2012}, who studied an earlier volume-limited sample, this finding is at odds with observations showing that at least 50\% of intermediate mass main-sequence stars (the progenitors of the WDs) reside in multiple systems. The explanation provided by \citet{Ferrario2012} was that the missing WDs may be concealed either in double degenerate systems or in Sirius-like systems. However, it now appears more likely that a large fraction of binaries merge during the post-main sequence phase. This topic will be covered extensively in \S\,\ref{origin}.

\subsection{Rotational Periods}\label{rot}

The spectroscopic determination of the rotational periods of WDs is very difficult because of their extremely pressure-broadened lines. However, the H$_{\alpha}$ absorption lines exhibit a very weak and narrow non-NLTE core that has allowed the studies of the rotational properties of hot ($\rm{T}_{eff}\gsimeq 14,000$\,K) DA WDs. Such studies have yielded null detections in the vast majority of WDs resulting in upper limits $\rm{v} \sin i \le 20\,\rm{km}\,{\rm s}^{-1}$ being placed on the projected rotational velocities \citep{Karl2005} which corresponds to lower limits of $P_{\rm rot} \ge 1$\,hr to the rotational periods.

Rotational periods can also be determined from the rotational splitting of the oscillation modes in pulsating WDs. Some well studied systems \citep[see, e.g.,][]{Winget1991, Obrien1996} have yielded rotation periods in the range $1.17-1.60$\,d corresponding to velocities of a few km\,s$^{-1}$. This may suggest a few days as the typical rotational period for non-magnetic WDs although significantly longer periods are much more difficult to establish without long term observations. The more recent studies of \citet{Hermes2017} of 27 pulsating hydrogen-rich WDs, the DAVs or ZZ Ceti stars, observed by the \emph{Kepler} mission \citep{Basri2005} provided constraints on the spin period of 20 of them, doubling the number of WDs whose spin periods are derived through asteroseismology. Follow-up spectroscopic observations, also conducted by \citet{Hermes2017}, provided WD's rotation as a function of mass. They find that WDs in the mass range $0.51-0.73$\,M$_\odot$, which are the progenies of main sequence stars in the mass range $1.7-3.0$\,M$_\odot$, have an average spin of $35\pm28$\,hr while the three most massive WDs in their sample seem to rotate much faster ($4.0\pm 3.5$\,hr). The measured values and upper limits are significantly smaller than the break-up speed. For instance, a 3\,M$_\odot$ main-sequence star that rotates uniformly with an equatorial velocity of $\sim200$\,km\,s$^{-1}$ would be expected to generate a WD with a rotation period of about $\sim 20$\,s which is not verified by observations. This is evidence of efficient transfer of angular momentum from the stellar core to the envelope during pre-WD evolution. Thus, core and envelope must be strongly coupled during the evolution with powerful stellar winds, particularly during the AGB evolution, carrying away most of the star's angular momentum \citep{Ferrario2005a}. The asteroseismology studies of about 300 red giant stars with data from the \emph{Kepler} mission have verified this hypothesis showing that degenerate cores already rotate more slowly than expected when they are at the end of the red giant branch evolution \citep{Mosser2012}. 

\begin{figure}[htbp]
\begin{center}
\includegraphics[width=0.8\textwidth]{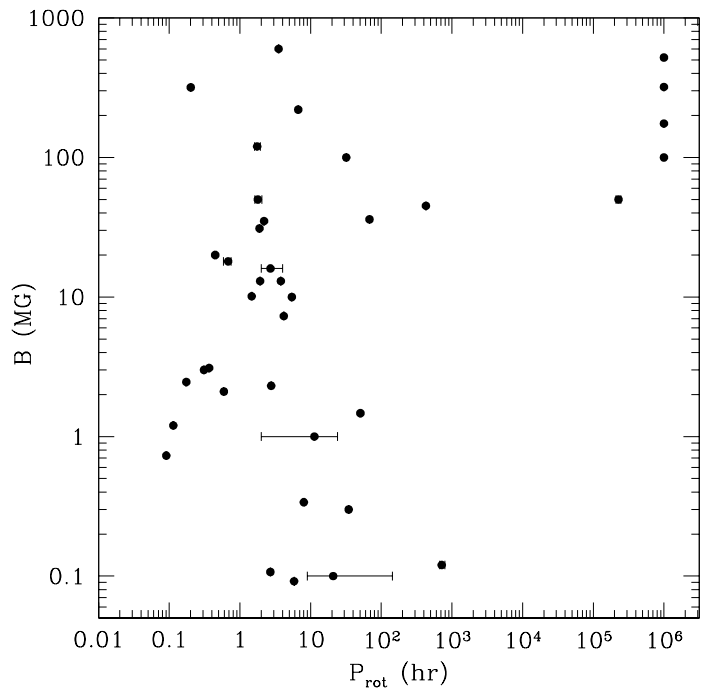}
\caption{The dependence of rotational period on magnetic field. Stars with estimated rotational periods of greater than 100\,yrs are plotted with a period of 100\,yrs.}
\label{rotation_distribution}
\end{center}
\end{figure}

In the case of the high-field MWDs, photometry and polarimetry can be used to determine or place more stringent limits on rotational velocities. Early studies of a few high-field MWDs showed no variations in polarisation over periods of tens of years. This was taken as evidence that they had periods  $\ge 100$ yrs.  The unlikely possibility that these stars may in fact have rotation periods that are significantly shorter than the typical integration times ($\sim$\,hrs) used in early studies was subsequently ruled out for Grw+70$^\circ$8247 and LP\,790-29 \citep{Jordan2002} by fast time resolution polarimetric studies. By using the changes in the polarisation position angle, which gives the rotation of the magnetic axis in the plane of the sky, \citet{Bagnulo2019} confirmed that the rotation period of Grw+70$^\circ$8247 is likely to be in the range $100- 1,000$\,years. Polarimetric variations have also been detected on a time-scale of $\ge 10$\,yrs in the high-field MWDs GD\,229 and G\,240-72 with implied rotational periods of $\sim 100$\,yrs \citep{Berdyugin1999}. The subsequent work of \citet{Brinkworth2013} confirmed the absence of short term variability in G\,240-72. Thus, it appears that there is a sub-class of high-field MWDs that is very slowly rotating which may be evidence that magnetic fields play an additional important role in the removal of angular momentum. The photometric studies of \citet{Brinkworth2013} of 21 high-field MWDs  revealed that 67\% are variable and 24\% have measurable rotation periods. They find no correlation between rotation and other WD parameters, such as mass, temperature, magnetic field, or age. Fig.\,\ref{rotation_distribution} shows the magnetic field versus rotation period of all currently known MWDs. Whilst we also find that there is no correlation between field strength and period, we also note that only strongly magnetic MWDs exhibit very long rotational periods.

There is also a significant class of rapidly rotating MWDs consisting of stars such as RE\,J0317-853 ($P_{\rm rot}=12$\,min) \citep{Barstow1995, Ferrario1997}. RE\,J0317-853 is believed from other considerations to be the result of a merger \citep{Ferrario1997, Vennes2003}.  We note that MWDs that are spun up in interacting binaries through the action of accretion torques could have typical rotation periods of hours.  Some of the isolated MWDs in the rapidly rotating group could therefore be mergers or the end products of the binary evolution of MCVs with very low-mass companions.

\subsection{The magnetic DQ white dwarfs}\label{DQ}

One of the most remarkable discovery in recent years is that of DQ WDs that are hot ($\sim 15\,000-24\,000$\,K) and whose atmospheres are carbon dominated ($-2\lsimeq\log[{\rm C}/{\rm He}]\lsimeq 2$). These objects show no H or He\,I lines in their optical spectra. Instead, they exhibit absorption lines of C\,II at 4267, 4300, 4370, 4860, 6578, and 6583\AA \citep{Dufour2005, Dufour2007, DufourHotDQ2008}. Interestingly, there are no DQ WDs with temperatures over 24,000\,K. Furthermore, approximately 70\% of them are magnetic \citep[][ and Kawka et al. (2020), in preparation]{Coutu2019, Dufour2013, Dunlap2015}.  \citet{Werner2012} have proposed that objects with CO dominated atmospheres \citep[such as H\,1504+65][which is extremely hot ($\sim 200,000$\,K), massive ($0.85\pm0.15$\,M$_\odot$), and shows no H or He lines in its spectrum]{Nousek1986}, could be the progenitors of the hot DQs.  \citet{Coutu2019} suggested that the massive ($M \ge 0.8\,M_\odot$) warm DQs ($10\,000 \le T_{\rm eff} \le 16\,000$ K) could be the descendants of the hot DQs. Most of these warm DQs are not known to be magnetic, but this could simply be due to the poor signal-to-noise of the spectra used in their analysis. High-resolution spectroscopy or spectropolarimetry with high signal-to-noise is necessary to determine the incidence of magnetism among warm DQs. 

\begin{figure}[htbp]
\begin{center}
\includegraphics[viewport=8 175 553 545,clip,width=8cm]{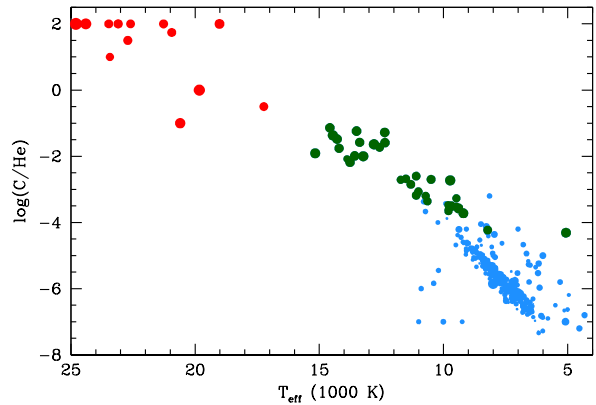}
\caption{Carbon abundances as a function of the effective temperature for known DQ stars. The point size is a function of the mass of the WD. Normal, low-mass DQs are depicted in blue. The higher mass warm DQs are in green and the hot and massive DQs are in red (Kawka et al., 2020a, in preparation).}
\label{C_Teff}
\end{center}
\end{figure}

\begin{figure}[htbp]
\begin{center}
\includegraphics[width=0.8\textwidth]{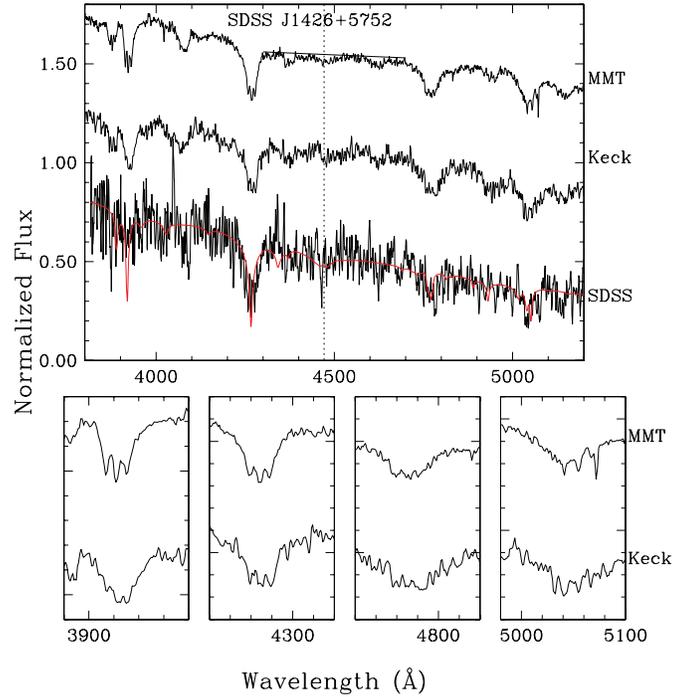}
\caption{Top panel: Spectra of SDSS\,J142625.71+575218.3 with overlapped in red the best, non-magnetic, fit \citep[see][for full details]{Dufour2008}. The dotted line indicates the position of the He\,I\,4471 line. Bottom panel: Carbon lines from the Keck and MMT spectra showing the presence of a 1.2\,MG magnetic field \citep[see][for full details]{Dufour2008}.}
\label{PulsatingMagneticDQ}
\end{center}
\end{figure}

A possible explanation for the origin of these carbon-dominated objects is that their AGB precursors could have undergone a final thermal pulse that resulted in the ejection of most, if not all, their helium envelopes. A reasoning along the same lines was proposed to explain the existence of hydrogen-poor WDs \citep[e.g.,][and references therein]{Werner2006}. However, this does not explain why so many hot DQs are also massive and magnetic. Two possibilities have been investigated: (i) they could be the progenies of the most massive main-sequence stars that can evolve into WDs ($5\lsimeq M/{\rm M_\odot}\lsimeq 8$), or (ii) they could be the result of stellar merging events, as proposed by \citet{Wickramasinghe14} and \citet{Briggs2015} to explain the origin of fields in all high-fields MWDs (see \S\,\ref{origin}). At present it is unclear how stellar merging occurs and whether it could lead to the ejection of most of the He layers of the merged object.  However, because the kinematic properties of the hot DQ WDs are those of an older stellar population \citep{Dunlap2015}, the stellar merging hypothesis seems to be by far the most likely cause for their origin \citep{Coutu2019}. This merging scenario is strongly supported by the recent study of \citep{Kawka2020b} of the ultra-massive warm DQ WD LP93-21 whose physical properties and highly retrograde halo kinematics can only be explained if this object is the outcome of an ancient double degenerate merger. We show in Fig.\,\ref{C_Teff} the carbon abundance as a function of the effective temperature for all known DQ WDs (hot and warm DQs plus also the cool DQ WDs described in section \ref{Carbon_lines}).
The temperature range of hot DQ WDs is such that some of them are expected to be pulsating \citep{Fontaine2008}. Hydrogen-rich WDs pulsate at temperatures ($T_{\rm eff}\approx12\,000$\,K) where H is partially ionized. Similarly, helium-rich WDs pulsate where helium is partially ionized at $T_{\rm eff} \approx 25\,000$\,K.  \citet{Fontaine2008} showed that the effective temperature of the small number of known hot DQs fall in the temperature range where carbon is partially ionized, and therefore should pulsate in non-radial g-modes similarly to WDs with hydrogen and helium dominated atmospheres. These pulsations can then be used to probe the interior structure of the WD. A search for pulsations in this class of objects led to the discovery by \citet{Montgomery2008} of the first pulsating, carbon-dominated WD, SDSS\,J142625.71+575218.3, which exhibits a period of 417\,s. Further observations of this intriguing WD conducted by \citet{Dufour2008} revealed that it is not only pulsating, but also possesses a 1.2\,MG field, as  the spectra of \citet{Dufour2008} depicted in Fig.\,\ref{PulsatingMagneticDQ} show.

Following the discovery of photometric variations in SDSS\,J142625.71+575218.3, several more hot DQ WDs have been found to be variable. Although this variability was originally attributed to pulsations, the lack of multi-mode variations, expected from pulsating WDs, and the discovery of a  2.1\,d photometric modulation in the hot DQ WD SDSS\,J000555.9-100213.5 \citep{Lawrie2013} led to the suggestion that the variations are not caused by pulsations but by rotation \citep{Williams2016}. This means that hot DQ WDs are generally characterised by rapid rotation which again supports the hypothesis that these objects originate from merging.

\subsection{The magnetic DAZ and DZ white dwarfs}\label{DZ}

\begin{figure}[htbp]
\begin{center}
\includegraphics[viewport=0 127 574 397,clip,width=8cm]{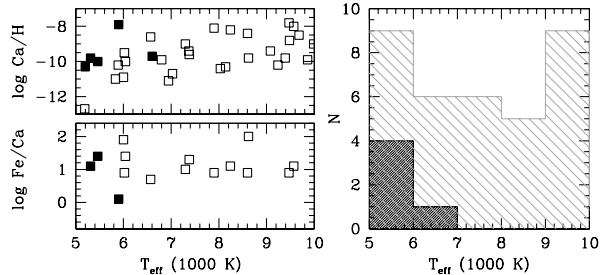}
\caption{Left: The abundances of cool DAZ WDs. The full squares represent the MWDs while the open squares the non-magnetic WDs. Right: The effective temperature distribution of DAZ WDs. The magnetic DAZ WDs are shaded in dark grey  \citep[from][]{Kawka2019_DZ}.}
\label{Kawka2019_cool_DAZ}
\end{center}
\end{figure}

As noted in \S\,\ref{Metal_lines}, about $25$\% of WDs show traces of metals in their spectra. For this metal pollution to exist it is necessary to assume ongoing accretion of planetary debris \citep{Jura2003} because the diffusion time-scales for these elements to sink is too short. This idea is strengthened by the discovery of gaseous and dusty discs around WDs \citep{Becklin2005, Jura2007, Farihi2008b, Farihi2009}. A subset of metal polluted WDs exhibits the presence of magnetic fields \citep{Reid2001, Kawka2011, Farihi2011, Kawka2014}. The studies of \citet{Kawka2014} have shown that the incidence of magnetism is higher in cool ($<6,000$\,K) and polluted hydrogen-rich WDs (denoted as DAZ) with fields measured to be in the range $\sim 70-500$\,kG \citep{Kawka2019_DZ}. The incidence drops to essentially zero at temperatures above about 7,000\,K. The studies of \citet{Kawka2019_DZ} of the current sample of DAZ WDs is summarised in Fig.\,\ref{Kawka2019_cool_DAZ} which clearly shows that the incidence of magnetism is close to 50\% at $T_{\rm eff}<7,000$\,K. The reason for this higher incidence of very weak fields at low temperatures is still unknown. \citet{Kawka2019_DZ} speculated that it could be caused by a gaseous planet plunging onto a WD. Such an event may instigate differential rotation over a large enough outer layer of the WD to generate the very weak fields detected in cool DAZ WDs. This mechanism was first suggested by \citet{Farihi2011} to explain G77-50. According to their picture, the orbits of outer planets and asteroids around the WD G77-50 could have been disrupted by a close encounter with another star. If, as estimated by \citet{Farihi2011}, such an encounter has a 50\% chance of happening every 0.5\,Gyr, then older WDs would be more likely to have accreted a gaseous giant planet during their motion around the Galactic centre than their younger counterparts.

Similarly, \citet{Hollands2015} showed that cool, helium-rich DZ WDs exhibit a high incidence of magnetism, i.e., about 13\% at $T_{\rm eff} \le 8000$\,K dropping rapidly at higher temperatures. Another peculiarity is that the strengths of magnetic fields in cool DZ WDs have a range from weak fields, comparable to the field strengths in cool DAZ WDs, to much stronger fields in excess of 10\,MG \citep{Hollands2015,Hollands2017}. It is possible that this is entirely a selection effect because the metal lines in DZ WDs are much stronger than in DAZ WDs.  Therefore at higher fields the weaker metal lines may be smeared out by magnetic field broadening (see \ref{fieldstruc}). The metal lines also become weaker at higher temperatures.

\subsection{Chromospheres, coronae and contemporary dynamos in WDs?}\label{dynamo1}

WDs develop convective envelopes as they cool, driven by the ionisation of HeI and HeII. It has been suggested that cool WDs with deep convective envelopes may generate fields of $\sim 100$\,kG driven by an $\alpha-\omega$ dynamo \citep{Thomas1995}. Such fields may be expected to lead to the formation of hot coronae with X-ray luminosities of $10^{27}-10^{30}$\,erg\,s$^{-1}$ \citep{Winget1994}. However, searches for coronal X-ray emission from cool WDs have so far led to negative results \citep[see][and references therein]{Musielak2003}. It is possible that in the case of WDs, the wave flux energy is dissipated in a chromosphere rather than a corona \citep{Musielak2005}. However, chromospheric activity is also not a general characteristic of MWDs.

From an observational point of view, there is no strong evidence to support that the incidence of magnetism increases in cooler WDs (but see \S\,\ref{fields}) as may be expected if a contemporary dynamo were in operation nor have there been reports on reversals in polarity in any MWDs.  Current observations are, however, limited to WDs with fields significantly larger than $10^3$\,G.

\subsection{Complexity of field structure, field decay and field evolution} \label{fields}

The free Ohmic-decay time can be approximated by $t_{\rm{ohm}}\sim \displaystyle{\frac{4\pi\sigma L^2}{c^2}}$ where $L$ is the magnetic field's variability scale-length and $\sigma$ is the electrical conductivity. If we take $L\sim R$ (WD's radius) and $\sigma$ consistent with what is expected in a WD's degenerate core, then $t_{\rm{ohm}}\sim 2-6\times 10^{11}$\,yr \citep{Cumming2002}. Calculations allowing for changes in conductivity with temperature in the outer layers of a WD yield decay time scales for the dipole ($l=0$) mode of $8-12\times 10^9$\,yrs and for the quadrupole ($l=2$) mode of $4-6 \times 10^9$\,yr for WDs masses in the range $0.6 - 1$\,M$_\odot$ and Carbon-Oxygen interiors \citep{Cumming2002, Wendell1987}. The non-linear coupling between the different modes that occurs when the Hall current is included could result in the growth of higher order modes relative to the dipole mode \citep{Muslimov1995}. However, there are estimates that have questioned whether this mechanism will give rise to the development of a dominant quadrupolar component in a typical cooling time \citep{Cumming2002}.

\begin{figure}[htbp]
\begin{center}
\includegraphics[width=0.8\textwidth]{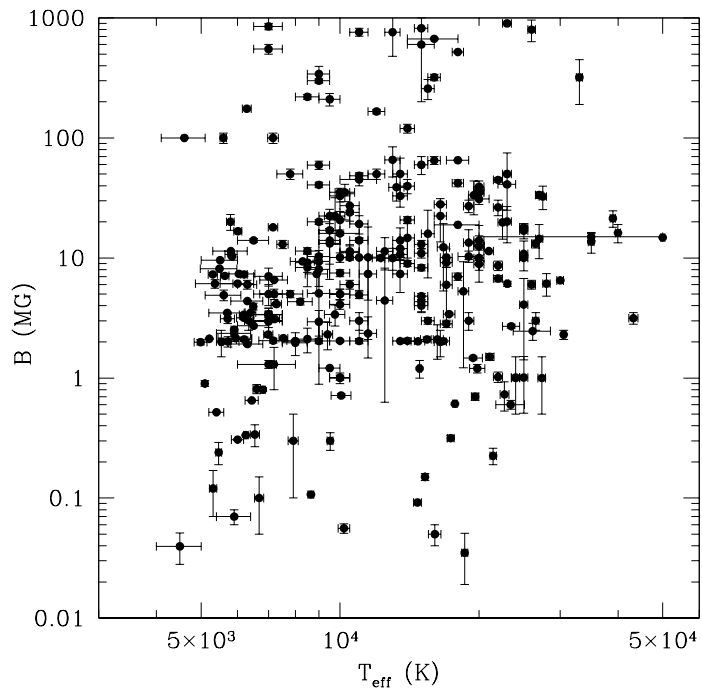}
\caption{Magnetic field strength against effective temperature showing no indication for field evolution with cooling age (this work).}
\label{temp_distribution}
\end{center}
\end{figure}

According to current observations (see Fig.\,\ref{temp_distribution}), there seems to be no evidence for a change in the incidence of magnetism or in field evolution as shown by \citet{Ferrario2015a}. That is, the mean field strength and distribution about this mean seem to be independent of effective temperature. They also show that the cumulative distribution function of the effective temperatures is smooth over the full range of temperatures, again suggesting that the incidence of magnetism has remained the same throughout the life of the Galaxy. The lack of evidence for the evolution of field strength with effective temperature is consistent with the long decay time scales outlined above.  However, the observational findings may not give us the full picture because the current sample of MWDs, by being Sloan-dominated, is very biased, favouring the detection of hot WDs with fields larger than a few MegaGauss. On the other hand, volume-limited samples such as that of \citet{Sion2014, Holberg2016}, contain too few MWDs to establish whether the incidence of magnetism is inversely proportional to effective temperature \citep[as first raised by][]{LiebertSion1979} or if there is a link between field strength and stellar age. Larger volume-limited samples comprising MWDs distributed over a wider range of field strengths and temperatures are needed to answer this long-standing question.

Definitive statements about field structure can only be made for a few well studied MWDs.  However, it is clear that complex field structures can be found in both the hotter \citep[e.g. REJ0317-853][]{Ferrario1997, Vennes2003} and the cooler \citep[e.g. WD1953-011][]{Maxted2000} MWDs. Furthermore, there are examples of both high field \citep{Euchner2005}, and low field \citep{Maxted2000, Landstreet2017} MWDs that appear to exhibit a similar level of field complexity. The complexity of the field could be linked to its origin. For instance, \citet{Garcia2012} suggested that MWDs resulting from mergers may have a more complex structure than those descending from the magnetic Ap and Bp main sequence stars \citep[see, e.g.,][]{Braithwaite2004}, if the conservation of magnetic flux from the main sequence to the compact star stage is a viable evolutionary path (see \ref{origin}).

\section{Magnetic interacting binaries}\label{binary}

High-field MWDs are the compact objects that accrete mass from a low-mass non-degenerate companion in the MCVs. These interacting magnetic binaries comprise the polars and the intermediate polars (IPs). The IPs are powered by Roche-lobe overflow and have stable and magnetically truncated accretion discs that mask the properties of the underlying MWD. In the strongly magnetic polars the coupling radius is larger than the circularisation radius so that accretion on to the MWD is not mediated by an accretion disc. Instead, matter is channelled onto the MWD's surface by magnetically confined flows \citep[e.g.,][]{Ferrario1999}. The presence of a sub-class of polars at first named ``low accretion rate polars (LARPs)'', where accretion occurs from the wind emanated by the companion star, was first recognised by \citet{Schmidt2005_LARP}. These objects were later identified as the possible progenitors of the MCVs by \citet{Tout2008} and re-named PREPs (pre-polars) by \citet{Schwope09} to avoid confusion (see \S\,\ref{origin}).
\begin{figure}[htbp]
\begin{center}
\includegraphics[width=0.7\textwidth]{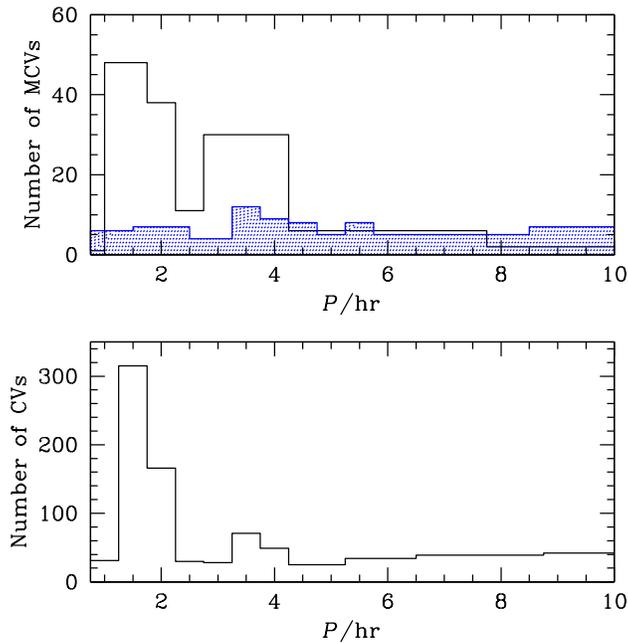}
\caption{Top: Orbital period distribution of MCVs. The polars are depicted with the black line histogram and the IPs with the shaded histogram. Bottom: Orbital period distribution of CVs taken from \citet{Ritter2003}. This figure is from \citet{Briggs2018b}).}
\label{MCV_Periods}
\end{center}
\end{figure}
In the polars the spin of the MWD is synchronised with the orbital period of the binary with the exception of a few systems, such as V1500\,Cyg, BY\,Cam, V1432\,Aql, V4633\,Sgr, CD\,Ind, Paloma, IGR\,J19552+0044, and RX\,J0838-2827. Generally, the two periods differ by less than 2\% but in two systems, Paloma \citep{Schwarz07} and IGR\,J19552+0044 \citep{Tovmassian2017} the desynchronisation is found to be somewhat larger ($\approx 3$\%). This asynchronicity has been attributed to recent nova explosions given that V1500\,Cyg, did undergo a nova eruption in 1975. Such systems are expected to reach synchronism over very short time-scales and \citet{Harrison2016} have reported that V1500\,Cyg has already become fully synchronized and exhibits an X-ray spectrum and luminosity that are consistent with those of a polar in a high state of accretion. 

\begin{figure}[htbp]
\begin{center}
\includegraphics[width=0.7\textwidth]{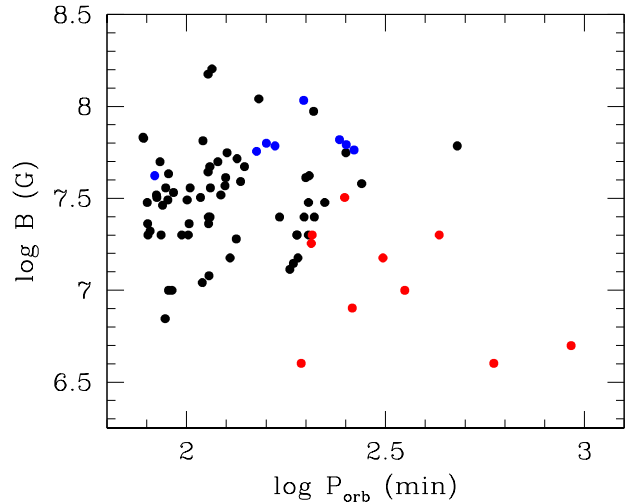}
\caption{Magnetic field strengths of MCVs against their orbital periods. The polars are in black, the PREPs in blue and the IPs in red.}
\label{MCV_Fields}
\end{center}
\end{figure}

The period distribution of the MCVs tends to be dominated by the polars at short orbital periods and the IPs at longer periods (see Fig.\,\ref{MCV_Periods}). Although the field strengths have been measured for only a handful of IPs, they generally seem to possess lower fields than the polars (see Fig.\,\ref{MCV_Fields}). It is however very likely that some of the most strongly magnetic IPs ($B\gsimeq 10^7$\,G) may eventually evolve into polars.  

Although the evolution of magnetic and non-magnetic CVs is expected to be similar, the calculations of \citet{Li1994} showed that magnetic braking is not as effective in strongly magnetic CVs because the magnetosphere of the MWD traps the companion star's wind. This hampers the angular momentum loss and mass transfer and leads to longer evolutionary timescales thus explaining why the fraction of MCVs among CVs is very high ($\sim 20-25$\,\%).

The polars exhibit high and low states of accretion and because they do not have a readily supply of material stored in an accretion disk, their shift from high to low states is quite sudden. During low states, the bare surface of the MWD can be observed and its magnetic field studied using the methods highlighted in \S\,\ref{zeeman}. None of the known IPs has ever been observed in a low state of accretion and this is one of the reasons why it is so difficult to determine the magnetic field distribution (and structure) of the MWDs in the IPs. The orbital periods of polars are typically in the range $P_{orb} \sim 80$\,min to $8$\,hrs and ideally suited for phase-resolved spectropolarimetric studies of the type already discussed for the isolated MWDs in \S\,\ref{fields}. The magnetic properties of the accreting MWD can also be studied from the cyclotron spectra arising from the accretion shocks that occur where the material that is channeled along field lines impacts on the stellar surface (usually close to the magnetic poles). This is the only method that can be used to estimate the fields of the most strongly magnetic and polarised IPs. In all other IPs the radiation emanating from the accretion shocks is hidden in the glare of the radiation emitted by their accretion disc and the region where the stream leaving the inner Lagrangian point impacts onto the disc (the so-called ``hot spot'').

\subsection{Field structure of the MWDs in MCVs}

In this section, we focus on our understanding of the magnetic field structure of the MWDs in accreting binaries.
\begin{figure}[htbp]
\begin{center}
\includegraphics[width=0.75\textwidth]{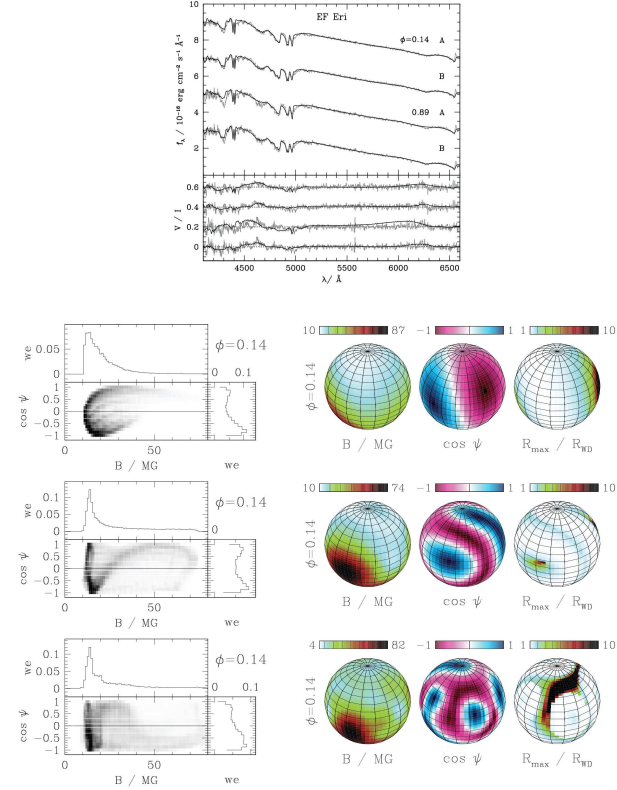}
\caption{Top: Flux and circular polarisation spectra of EF\,Eri at two orbital phases for an offset dipole model (marked A) and a multipole expansion model (marked B). The black curves are the best fit models superimposed to the observed spectra in grey \citep[see][for further details]{Beuermann2007}. Bottom: The $B-\psi$ diagrams, where $\psi$ is the angle between the local field direction and the line of sight, corresponding to an offset dipole (top), the full multipole expansion up to $l_{\rm max}=3$ (centre), and the full multipole expansion up to $l_{\rm max}=5$ (bottom). The weight distributions (we) are also shown. The right panels show the inferred distributions of the field strength $B$, $\cos\psi=B_l/B$ (where $B_l$ is the field component along the line of sight) and of the maximum radius $R_{\rm max}$ to which the field lines extend \citep[see][for further details]{Beuermann2007}}
\label{EF_Eri}
\end{center}
\end{figure}
The MWDs in the three polars EF\,Eri, BL\,Hyi and CP\,Tuc that have been extensively studied using Zeeman tomography \citep[][see Fig.\,\ref{EF_Eri}]{Beuermann2007}, exhibit field structures that are at least as complex as those in the two isolated high-field MWDs discussed in \S\,\ref{fieldstruc} using the same method. The additional constraints provided by the locations of the accretion shocks are also generally consistent with field structures that have strong contributions from multipoles higher than the dipole (see below). The study of the polarised radiation arising from the accretion shocks provides an excellent method to investigate the field strength and structure of the MWD in the MCVs. In these systems, the material that is channelled from the orbital plane to the surface of the MWD forms stand-off shocks that achieve temperatures 
\[
T=3.7\times 10^8\left(\frac{M_{\rm wd}}{M_\odot}\right)\left(\frac{10^9}{R_{\rm wd}}\right)~{\rm K}. 
\]
The shocked material emits bremsstrahlung radiation in the hard X-rays and cyclotron radiation in the UV to IR spectral regions (depending on field strength). At high accretion rates and/or low fields the shock is mostly bremsstrahlung dominated. At low accretion rates and/or high fields the main emission mechanism is thermal cyclotron caused by electrons gyrating in the WD's magnetic field. The observed soft X-ray component arises from the heated photosphere of the WD near the accretion shocks \citep{lamb_masters79}. However, some polars appear to display some puzzling anomaly whereby their soft X-ray luminosity, $L_{\rm soft}$, is much larger than expected from simple photospheric heating caused by the re-processing of the two primary, $L_{\rm brems}$ and $L_{\rm cycl}$, sources of radiation. That is, $L_{\rm soft}>> L_{\rm brems}+L_{\rm cycl}$. \citet{kuijpers_pringle82} suggested that dense clumps of material penetrate into the photosphere of the WD to large optical depths so that the shock becomes buried. Under this condition the energy is emitted as black body radiation at a temperature much larger than the effective temperature of the WD. 
\begin{figure}[htbp]
\begin{center}
\includegraphics[width=0.7\textwidth]{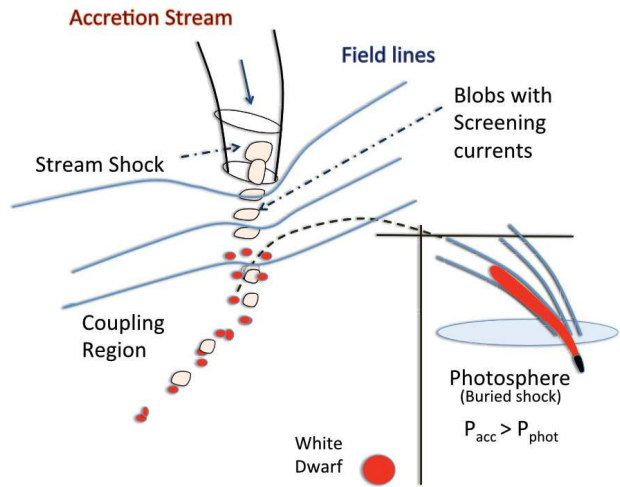}
\caption{Schematic diagram explaining scenario 1 (see text). The shock generated at the region of impact of the stream on to the MWD's magnetosphere breaks up the material into different mass blobs because of the Rayleigh-Taylor instability limited by the strength of the magnetic field. The more massive blobs follow a dynamical trajectory but are gradually stripped of matter by the Kelvin-Helmholtz instability. The clumps get pulled and form discrete filaments as they free fall along field lines. At high specific accretion rates ($\gsimeq 10^2$\,g\,cm$^{-2}$) the resulting shock is buried in the MWD's atmosphere and the accretion energy is released as a reprocessed soft-X-ray black-body component \citep[this figure is from][]{Wickramasinghe2014_Acc}.}
\label{blobby}
\end{center}
\end{figure}
It has become apparent through the study of accretion processes in polars that MHD and other instabilities occurring at the magnetospheric boundary, and that are not usually encapsulated in large scale MHD simulations of accretion flows, determine the nature of field channelled flow onto the MWD. 

Two possible closely related scenarios may explain why the accretion in some polars should be clump-dominated (as inferred from their soft X-ray excess) whilst in others it is not. In the first scenario (scenario 1) a strong shock forms at the stream-magnetosphere boundary. Various interchange instabilities, such as the magnetic Rayleigh-Taylor instability in the cooled post-shock region, result in the stream fragmenting into clumps of gas with length scales and masses that are set by the strength of the field. As this material penetrates the magnetosphere the initially less massive clumps couple on to field lines and feed the main accretion pole (that is, the stronger accretor) while the more massive clumps penetrate further down stream and are shredded into smaller blobs by the Kelvin-Helmholtz instability which then couple on to field lines and feed the second pole (the weaker accretor). This model is consistent with observations of systems for which there is direct evidence for X-ray emission at $\sim 0.5$\,keV from a strong stream shock but perhaps not for other systems where evidence for such emission has not been reported. We show in Fig.\,\ref{blobby} a schematic diagram depicting this scenario.

In an alternative but closely related scenario (scenario 2) the stream is continuously decelerated by a series of weak shocks as it plunges into the magnetosphere. MHD and other instabilities at its surface shred the stream into small enough mass components that diffuse into co-rotating field lines and deplete the mass of the stream until it is exhausted. This scenario is similar to what may be in operation in the inner transition region in an accretion disc where material is similarly depleted from its surface layers until the disc material is exhausted. However to date there has been no reported evidence for blobby accretion in the IPs. The physics of the stream-magnetosphere interaction has been described in more detail in the review of \citet{Wickramasinghe2014_Acc}. 
\begin{figure}[htbp]
\begin{center}
\includegraphics[width=0.45\textwidth]{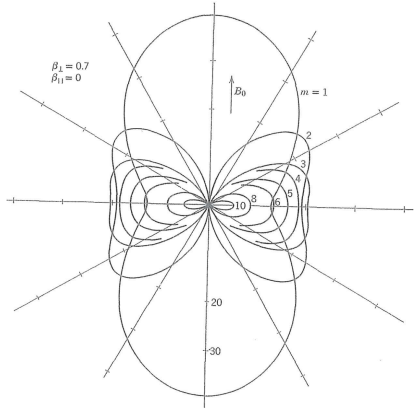}
\caption{Angular distribution of cyclotron radiation from a mildly relativistic electron \citep[after][]{Oster1960}. Here $B_0$ is the magnetic field direction and $m$ the cyclotron harmonic number. The curves labelled  $m=1, 2, 3\cdots$ enclose the volume of the beam of radiation emitted by the electron and show that as the harmonic number increases, the radiation is concentrated to an increasingly narrower cone whose axis is perpendicular to the direction of the magnetic field.}
\label{Oster}
\end{center}
\end{figure}

\begin{figure}[htbp]
\begin{center}
\includegraphics[width=0.5\textwidth]{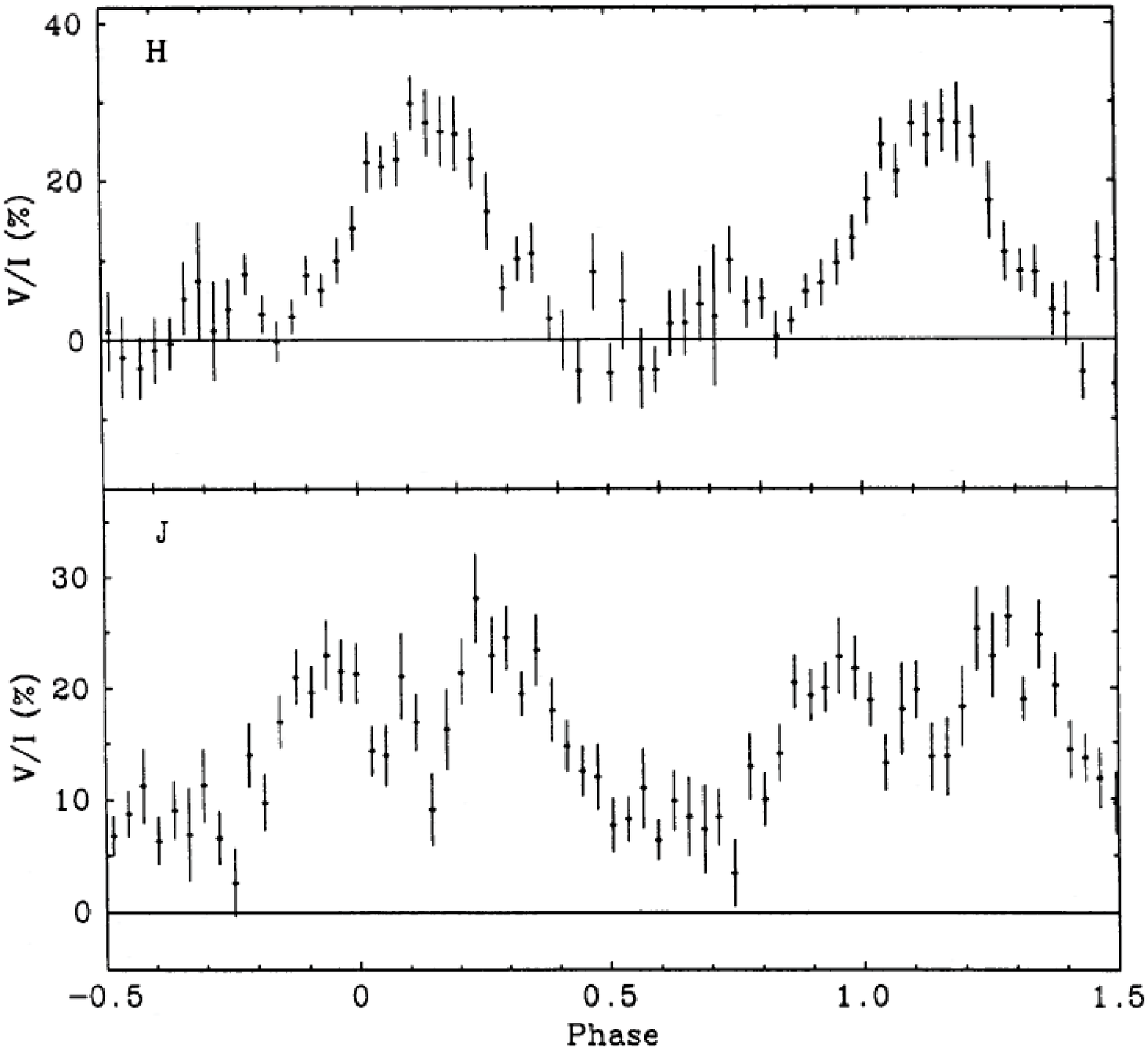}
\caption{Circular polarisation curves of V834\,Cen in the H and J infrared bands \citep[figure from][]{Ferrario92}}.
\label{V834Cen}
\end{center}
\end{figure}

\begin{figure}[htbp]
\begin{center}
\includegraphics[width=0.98\textwidth]{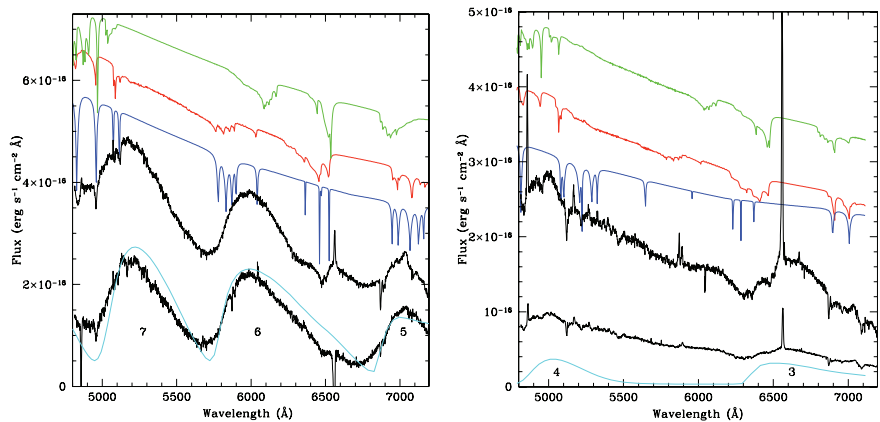}
\caption{Left panel: Observed spectra of VV\,Pup (black curves). The upper spectrum is the average of all bright phase spectra of VV\,Pup while the lower spectrum is obtained by subtracting the average faint phase spectra from the average bright phase spectra.  The cyclotron model (cyan line) is overlapped to the spectrum at the bottom and the numbers denote the cyclotron harmonics which correspond to a field of 31.5\,MG. Right panel: Two faint phase spectra of VV\,Pup (black lines) taken at two different epochs with overlapped the cyclotron model (cyan line) which correspond to a field of 54.6\,MG (the numbers denote the cyclotron harmonics). Halo Zeeman features (blue lines), and photospheric Zeeman models (off-centre dipole in green and centred dipole in red) are also shown at the top of each panel. All details of observations and modelling can be found in \citet{Mason2007}.}
\label{VVPup_Fig}
\end{center}
\end{figure}

\begin{figure}[htbp]
\begin{center}
\includegraphics[width=0.5\textwidth]{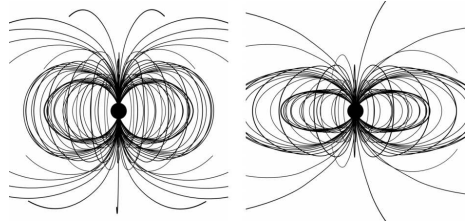}
\caption{Schematic diagram of a dipole (left) and an advected dipole (right).}
\label{Advected_dipole}
\end{center}
\end{figure}

Cyclotron radiation is strongly polarised and beamed orthogonally to the magnetic field direction at harmonics numbers $m>1$ as shown in Fig.\,\ref{Oster}. If the line of sight of a distant observer forms an angle $i$ to the spin axis of the MWD and the dipole magnetic axis is inclined by an angle $\beta$ to the rotation axis, then as the MWD rotates the angle $\psi$ between the line of sight and the field direction at the emission region varies as a function of the orbital phase $\phi$. Because cyclotron radiation is beamed, the observer will see strongly modulated intensity and polarisation curves \citep{Meggit_Wickramasinghe82}. All polars, together with a handful of IPs, exhibit a very high level of circular polarisation. As an example, we show in Fig.\,\ref{V834Cen} the circular polarisation curves in the H and J bands of the polar V834\,Cen.
Because thermal broadening dominates at most viewing angles, cyclotron radiation generally appears as continuum emission. Nonetheless, if the shock's electron temperature is sufficiently low to prevent relativistic mass broadening to dominate, then resolvable cyclotron emission lines may become visible at those orbital phases when the magnetic field lines are nearly orthogonal to the line of sight \citep{Meggit_Wickramasinghe82}. If the accretion rate is high enough, the magnetically channelled material can flow to both foot points of closed field lines (that is, not only to the most favourably inclined pole). Thus it may be possible to detect two sets of cyclotron lines and the separation of consecutive harmonic bumps provides us with a precise measure of the field strength at both shocks (above and below the orbital plane). At low temperatures the position of the m$^{\rm th}$ harmonic for a magnetic field $B$ and viewing angle $\psi = 90^\circ$ to the magnetic field direction is given by:
\begin{equation}
\lambda_m=\frac{10\,710}{m}\left(\frac{10^8\,{\rm G}}{B}\right)
\end{equation}
Cyclotron spectroscopy has proved to be a very powerful method to gain information on the magnetic field strength and structure of the MWD in MCVs as first shown by \citet{Visvanathan1979}. As an example, we show in Fig.\ref{VVPup_Fig} some spectra of VV\,Puppis from \citet{Mason2007} exhibiting two sets of prominent cyclotron harmonic features showing that the MWD was accreting near both magnetic poles. 
\begin{figure}[htbp]
\begin{center}
\includegraphics[width=0.5\textwidth]{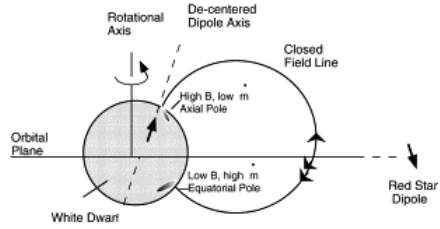}
\caption{Schematic diagram of a MWD that accretes matter on to two regions. The more favourably inclined, and thus more strongly accreting region, is that located below the orbital plane. This region has a weaker magnetic field than the second located above the orbital plane.}
\label{Dipole-offcentred}
\end{center}
\end{figure}
During the bright orbital phase the spectra show the presence of a field of 31.5\,MG (harmonic numbers 5, 6, and 7) while during the faint phase a second set of harmonics becomes visible revealing a much stronger field of 54.6\,MG (harmonic numbers 3 and 4). The Zeeman absorption features, also present in the spectra of VV\,Pup, have been modelled as arising either from an accretion halo surrounding the cyclotron emission regions or from the photosphere of the MWD. Neither of these explanations has been satisfactory thus indicating that the field geometry of VV\,Pup is likely to be much more complex than can be modelled with a simple centred or off-centred dipole, as in the case for EF\,Eri \citep[see above and][]{Ferrario1996}. We show in Fig.\,\ref{Dipole-offcentred} a schematic diagram of a MWD with an off-centred dipole that is accreting material on to two regions. In this figure the most favourably oriented region is that below the orbital plane. To date, we have been able to determine the field strengths of most polars through either the modelling of Zeeman split lines, or cyclotron spectroscopy, or in some cases both. Cyclotron harmonics have been detected in the IR \citep[e.g.,][]{Bailey91, Ferrario93}, in the optical \citep{Visvanathan1979, SchwopeBeuermann1990, Ferrario94, Mason2007} and in the most strongly magnetic systems also in the UV \citep[e.g.,][]{rosenetal01-1,Gaensicke01}. 
It is possible that accretion may modify the field strength and/or structure through accretion induced field decay \citep{Cumming2002}. The calculations of \citet{Zhang2009} indicate that magnetically confined flows that impact on to the surface of the MWD near the foot points of closed field lines, can advect the field toward the equator (see Fig\,\ref{Advected_dipole}) where it becomes buried.
This leads to a reduction in the polar field strength down to a bottom field of just a few $10^3$\,G, regardless of the initial field strength. This mechanism was first invoked by \citet{Zhang2006} to explain the low fields of the millisecond radio pulsars that have undergone recycling via accretion from a companion star.  However, if the total mass accreted is less than about $0.1-0.15$\,M$_\odot$, then the field diffuses faster than can be advected and is not modified by accretion. The mass accretion rates in polars, generally less than about $10^{16}$\,g\,s$^{-1}$, is low enough to prevent field restructuring. In the IPs, however, the much higher accretion rates could in principle cause significant field restructuring and decay. On the other hand, mass-loss caused by nova explosions (not present in neutron stars) could hinder the accumulation of mass onto the MWD and thus not even highly accreting IPs may suffer from field restructuring decay. Systems that have reached the bottom field are predicted to exhibit a substantially non-dipolar magnetic field structure and could be found among those dwarf novae with quasi-periodic oscillations \citep{Zhang2009}. Field screening in the IPs during phases of accretion from a disc has also been proposed by \citet{Webbink2002} and \citet{Cumming2002}.

Accretion in MCVs occurs via magnetically confined funnels in polars and accretion disc plus magnetically confined curtains in IPs  \citep[see, e.g.,][]{Ferrarioetal93}. However, the IPs FO\,Aqr, TX\,Col, BG\,CMi, AO\,Psc, and V1223\,Sgr exhibit characteristics of both types, that is, while most of the accretion flow is mediated by an accretion disc, part of the flow appears to run over the disc to thread directly onto the magnetic field lines (e.g., see the work of \citet{Hellier1993} on the IP FO\,Aqr). The IP V2400\,Oph is, instead, a disc-less accretor \citep{buckleyetal95}. As the star rotates around its axis a magnetically confined funnel impacts first near one of the two magnetic poles and then near the other \citep[see][]{Hellier91, FerrarioWick99}. The MCV Paloma has been proposed by \citet{Joshi2016} to also be a disc-less IP similar to V2400\,Oph rather than an asynchronous polar.

Because of the radiation properties of the MCVs, the discovery of an unresolved hard (20-40\,keV) X-ray emission in the inner 10\,pc Galactic Centre by the Nuclear Spectroscopic Telescope Array \citep{Perez2015} has been attributed to a large population of IPs \citep{Pretorius2014,Hailey2016}. Non-magnetic CVs and polars, instead, may be responsible for the much softer X-ray emission originating from the Galactic ridge \citep{Xu2016}.

\section{Origin of magnetic fields in isolated and accreting WDs}\label{origin}

There are two main theories for the origin of magnetic fields in WDs. The first is that the fields are essentially of fossil origin reflecting in some way the magnetic flux of their progenitor stars dating back to at least the main-sequence \citep{Woltjer1964, Mestel1966, Moss2003, Tout2004}. The second is that fields are generated in convective dynamos when two stars, one of which with a degenerate core, merge \citep[see the various merging scenarios proposed by][]{Tout2008, Nordhaus2011, Garcia2012, Wick2014}. In the following sections we will review the work done on the origin of magnetic fields in WDs and the pros and cons of the two different generation theories noting that a full review on field generation in all stars can be found in \citet{Ferrario2015b}.

\subsection{Fossil fields}\label{fossil}

The observed WDs are the end products of the evolution of main-sequence stars of mass $\sim 1.1 - 8$\,M$_\odot$. In this mass range, ordered large-scale magnetic fields of $\sim 300 - 30 $\,kG are observed in about 10\% of early type stars \citep[e.g.,][]{Wade2016}.  Their magnetic fluxes are similar to those of the MWDs at least at the high end of the field distribution suggesting a possible evolutionary link between these two groups of stars.

Additional support for a fossil origin of the fields in the high-field MWDs comes from their average mass being higher than that of their non-magnetic counterparts (see \S\,\ref{masses}). This indicates that if high-field MWDs are the result of single star evolution they must, on average, descend from progenitors that are more massive ($\ge 2$\,M$_\odot$) than those of non-magnetic WDs. The masses of the main-sequence Ap/Bp stars that host strong, large-scale magnetic fields would fulfill this requirement.

There are strong arguments to suggest that fields in the magnetic main-sequence stars are not generated by a contemporary dynamo but are of fossil origin. The discovery of global organised magnetic fields in about $7\%$ of the pre-main sequence Herbig Ae/Be stars ($M\sim 2-15$\,M$_\odot$) by \citet{Alecian2008} lends further support to this view. Calculations of pre-main-sequence stellar evolution have also shown that stars more massive than $\sim 2$\,M$_\odot$ are expected to reach the main sequence retaining a radiative core \citep{Palla1993} in  which a magnetic field may be preserved.
 
The numerical studies of \citet{Braithwaite2004} have shown that the stability of fossil fields in Ap/Bp stars require linked poloidal-toroidal field structures. Their toroidal component resides inside the star while the external field is approximately dipolar with an offset from the centre of the star much like the fields observed in Ap/Bp stars. The subsequent evolution of such fields remains largely unexplored although various scenarios have been discussed \cite[e.g.,][]{Tout2004,Quentin2018}. Numerical magneto-hydrodynamical calculations have shown that while the dynamo mechanism is effective in generating weak fields of the order of the equipartition value in the convective cores of stars with radiative envelopes \citep{Brun2005}, much stronger fields can be generated if a fossil field is present in the radiative region \citep{Featherstone2009}. It is possible that the interplay between toroidal and fossil poloidal components and the field generated in the convective core during stellar evolution may lead to a MWD's surface field structure that is much more complex than in the progenitor main-sequence star.

\begin{figure}[htbp]
\centering
\includegraphics[width=0.5\columnwidth]{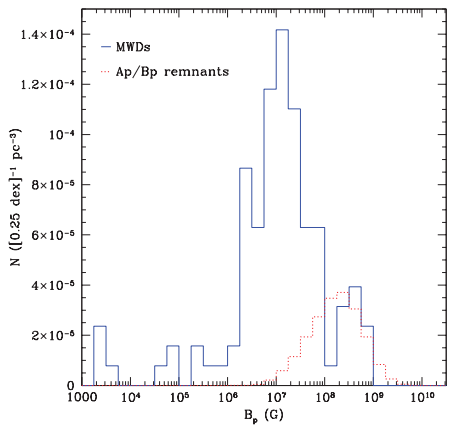}
\caption{Solid line: incidence of magnetism among MWDs. Dotted line: predicted incidence of the remnants of magnetic Ap/Bp main-sequence stars \citep[from][]{Kawka2004}.}
\label{MS_fossil}
\end{figure}

An argument against the fossil field hypothesis is that the birth rates of the magnetic main-sequence Ap/Bp stars are not consistent with those of the MWDs as first noted by \citet{Kawka2004} (see Fig.\,\ref{MS_fossil}). A better agreement can be achieved if it is assumed that in addition to the observed Ap/Bp stars, $40$\% of main-sequence stars more massive than $4.5$\,M$_\odot$ have fields in the  $10-100$\,G range and also evolve into high-field MWDs \citep[see][and their Figures\,5 to 8 for the various proposed scenarios]{Wickramasinghe2005}. However the ensuing  spectropolarimetric surveys of \citet{Auriere2007} all but excluded the existence of these weak fields, revealing, instead, a ``magnetic desert'' below 300\,G.

The arguments for and against a fossil origin of fields from single star evolution should be weighed against an important piece of evidence involving the seemingly lack of binaries consisting of a MWD with a non-degenerate and fully detached companion star \citep{Liebert2005, Liebert2015}.

In a recent paper, \citet{Quentin2018} follow the evolution of a 3\,M$_\odot$ star with fossil toroidal and poloidal fields through the various stages of stellar evolution up to the point when a degenerate core is formed. Their calculations use a 1-D stellar evolution code modified to allow for differential rotation and use averaged large-scale poloidal and toroidal magnetic fields that evolve following the $\Omega$ and $\alpha-\Omega$ dynamo mechanisms. The models follow variations on nuclear time scales and are not equipped to allow for dynamo cycles as for instance in \citet{Brun2005} and \citet{Featherstone2009}. They confirm a previous proposal by \citet{Tout2004} that large scale average fields are attenuated in the convective regions compared to radiative regions. The calculations show that a 3\,M$_\odot$ star evolves into a MWD with strong toroidal fields of $10^6-10^8$\,G concentrated in its outer layers but with a very much weaker poloidal field almost independently of initial conditions.  This suggests that all 3\,M$_\odot$ stars (and by extrapolation all $3-8$\,M$_\odot$ stars) are likely to end their evolution with weak poloidal fields that will place them in the very low-field regime. The authors argue that the lower mass ($<3$\,M$_\odot$) main-sequence stars with extensive convective envelopes are likely to evolve into non-magnetic WDs. The possibility that fields may be generated in stars during post main-sequence evolution in a core-envelope dynamo during helium-shell burning and later be encapsulated in the WDs resulting in low-field MWDs ($B\lsimeq 10^5$\,G) was already proposed by, e.g., \citet{Levy1974}.

\subsection{High-field MWDs from common envelope mergers}\label{mergers}

\begin{figure}[htbp]
\centering
\includegraphics[width=0.7\textwidth]{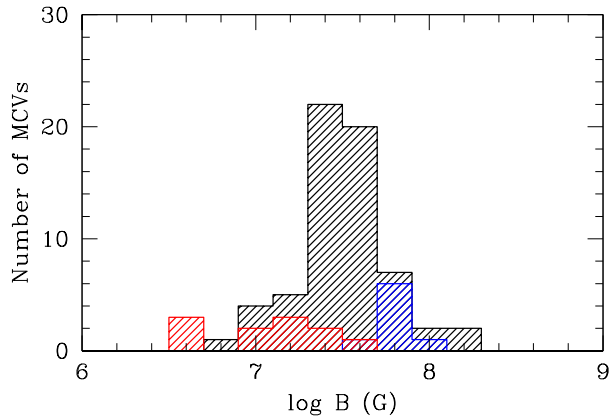}
\caption{Magnetic field distribution of polars (black), IPs (red), and PREPs (blue).}
\label{MCV_Bfield_histo}
\end{figure}

It was first noted by \citet{Liebert2005} that none of the $\sim 500$ spectroscopically observed WD+M pairs in the Sloan Digital Sky Survey (SDSS) with spectral resolutions that can measure fields greater than a few $\sim 10^6$\,G is magnetic. A follow-up study of a much larger sample of SDSS objects  ($\sim 1,700$) led to the same conclusion \citep{Liebert2015}. Likewise, of the $\sim 200$ MWDs with $B \ge 10^6$\,G that have so far been catalogued \citep{Ferrario2015a}, none has been found to be paired with a detached non-degenerate companion. However, we know that MWDs are commonly paired with such stars and accrete mass from them, either through Roche lobe overflow or stellar wind (see \S\,\ref{binary}). These systems are the MCVs. Therefore, the intriguing aspect of this finding is that there are no wide binaries consisting of a MWD paired with a non-degenerate and non-interacting companion star  even if wide-orbit pairs are very common among non-magnetic WDs \citep[$\sim 25$\%, see ][]{Rebassa13}.  This led to the suggestion that a significant proportion of the known isolated MWDs (if not all of them) may be the end product of close binary evolution as first suggested by \citet{Tout2008}.
\begin{figure}[htbp]
\centering
\includegraphics[width=0.7\textwidth]{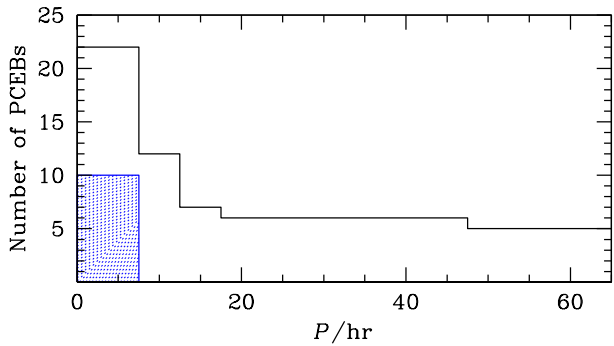}
\caption{The orbital period distributions of PCEBs  \citep[black solid line,][]{nebot13} and PREPs (shaded blue histogram).}
\label{PCEB}
\end{figure}
The basic idea is that magnetic fields can be generated in differentially rotating common envelopes (CE) and become frozen into the WDs. The strongest fields are expected in WDs that are formed when two stars merge during a CE phase of evolution \citep{Wickramasinghe2014}. The end product would be an isolated high-field MWD with a higher than average mass. Binaries whose components survive CE evolution and with their cores coming close enough to each other to generate a strong field in the WD, will evolve first into wind-accreting PREPs and then into bona-fide MCVs. The short orbital periods of the PREPs as compared to post-common envelope binaries (PCEBs, see Fig.\,\ref{PCEB}) and the high magnetic fields of PREPs (see Fig.\,\ref{MCV_Bfield_histo}) seem to support this hypothesis. If the stars emerge from CE with a large separation, they may either evolve into non-magnetic CVs or may never come into contact. There are some double degenerate binaries (DDs) consisting of a MWD with a non-magnetic WD companion \citep[a list of these DDs can be found in][]{Kawka2017}. These could be the outcome of an initially triple system where two stars merged and generated a MWD, while the third component evolved in isolation to become a non-magnetic WD. The CE stellar interaction/merging hypothesis of \citet{Tout2008} 
and \citet{Wickramasinghe2014} is supported by the population synthesis calculations of \citet{Briggs2015} and \citet{Briggs2018a}, in relation to the high-field MWDs, and \citet{Briggs2018b}, in relation to the magnetic fields of the WDs in the MCVs \citep[but see also the work of][]{Belloni2019}. The calculations of \citet{Briggs2015} have shown that the incidence of magnetism among WDs together with their  mass distribution are consistent with the view that they are the outcome of stellar merging during common envelope evolution.  A follow-up study by \citet{Briggs2018a} has further demonstrated that the magnetic field strength and distribution of high-field MWDs are also consistent with the merging hypothesis. These studies were carried out by generating a population of binaries on the main-sequence and by evolving them to the age of the Galactic disc using the rapid binary stellar evolution algorithm (BSE) of \citet{Hurley2002}. The  conditions they set are that (i) as the stars enter the common envelope stage the more massive of the two (the primary) has a degenerate core in which the newly formed magnetic field can freeze-in and (ii) no further nuclear burning in the core of the star that will evolve into a WD can take place because such burning would destroy any frozen-in magnetic field. If one of these two conditions is not met the stars will evolve into non-magnetic isolated WDs or, if they have a companion, into non-magnetic binaries. In their calculations they also include double WD binaries that merge to form a single WD at any time after their last common envelope evolution. The hot DQ WDs (see section \ref{DQ}) have been proposed to be the single WD outcome of the merging of two WDs.

\section{Summary and outlook}\label{conclusions}

The magnetic fields of isolated and accreting MWDs range from $\sim 10^{3}$ to $\sim 10^{9}$\,G with a field distribution that peaks near a few $\times 10^7$\,G.  While the high-field cut-off is expected to be real, studies of the low-field distribution are compromised by the sensitivity of current spectropolarimetric surveys in the case of isolated MWDs and by the inability to observe the surface of the naked MWD in accreting binaries. In the case of accreting binaries, field strengths are either indirectly estimated through the studies of the observed accretion characteristics or, in the most strongly magnetic systems, through the detection of Zeeman and/or cyclotron lines from the MWD's surface.

The isolated MWDs appear to divide into two groups. A high-field group ($10^5 - 10^9$\,G) which dominates the current statistics because they are the easiest to discover, and a low-field group ($\lsimeq 10^5$\,G), the importance of which has become apparent from spectropolarimetric surveys using 8\,m class telescopes. Population studies of the $20$\,pc volume-limited sample of MWDs suggest that at least $13$\% of all WDs are magnetic but the sample is incomplete at the kilo-Gauss level. Future spectropolarimetric surveys of even higher sensitivity may reveal that the incidence of magnetism increases at the very low field end. It is thus possible that all WDs possess a global, albeit weak, dipolar magnetic field. If this turns out to be the case, the accretion onto the WD in CVs may always be mediated, at least very close to the stellar surface, by the WD's magnetic field.

The tomography study of the field structure of a few isolated and accreting high-field MWDs have confirmed the presence of quite complex field structures. Other studies have indicated that complexity in field structure is not restricted to the very high field stars. Regardless of field complexity, there is so far no evidence that magnetic fields evolve with cooling age. This is consistent with theoretical estimates of the decay time-scales of the lower order dipole and quadrupole modes (4-10 Gyrs) although a larger and unbiased sample of MWDs (currently our sample is dominated by objects discovered by the SDSS) is critical to finally establish whether the incidence of magnetism is higher among cool WDs.

WDs tend in general to be slow rotators, which is evidence of efficient transfer of angular momentum from the stellar core to the envelope during pre-WD evolution (see \S\,\ref{rot}). Several MWDs, however, exhibit anomalously slow rotation periods with some estimated to have periods in excess of $100$\,yrs. The MWDs in this group have high fields suggesting that this slow rotation may be a result of enhanced braking of the stellar core due to the presence of the magnetic field. There is also a group of high-field MWDs with a mean rotation period of a few days which may be typical for normal (non-magnetic) WDs as derived from the studies of their pulsation modes. However, some MWDs have rotation periods of less than a few hours and these could be the outcome of stellar merging or could be the end products of MCV evolution.

The significant increase in the number of known isolated and accreting MWDs has led to new insights on the nature of magnetism in these stars. However, several outstanding problems will need to be addressed before major advances can be made in this area of research. These include, for instance, the construction of more realistic model atmospheres that allow for magnetic fields. Further insights into the field structure of MWDs will require the use of more sophisticated atmospheric models combined with Zeeman tomographic observations extended to include all four Stokes parameters.

We conclude by noting that recent observations have shown that it is now possible to gain insights on planetary systems through the studies of the metal-polluted atmospheres of many isolated WDs. This opens up the intriguing possibility of the detection of Earth-type planets around MWDs through the discovery of the anomalous atmospheric heating or radio emission predicted by the unipolar inductor model (as it has been suggested to explain the emission lines observed in MWD GD\,356), or of gaseous Jupiter-like planets through the detection of cyclotron emission from wind-driven accretion onto the MWD. The study of MCVs has also revealed the possible first discovery of a giant planet orbiting one of these systems. Thus planets and planetary systems associated to isolated and post-CE binaries may be more widespread than ever anticipated and can provide us with important clues on the formation of first and second generation planets.

Other studies have demonstrated that the polarised emission from the photospheres of MWDs could be used to place limits on fundamental physics, such as the coupling constants in some non-standard theories of gravity by measuring gravitational bi-refringence. 

\section*{Acknowledgements}

The authors thank St\'ephane Vennes for useful discussions.

\section*{References}

\bibliography{cospar}

\section*{Appendix}

Table\,\ref{tbl-split1} lists three different angular momenta, L\'ande factors and relative intensities of the Zeeman components for the most commonly detected element lines in WD atmospheres. The L\'ande factors and relative intensities were calculated using the equations in \S\,\ref{Metal_lines}.

\begin{table}[t!]
\caption{L\'ande factors and relative intensities of the most detectable metals.} \label{tbl-split1}
\centering
\begin{tabular}{ccrcccrc}
\hline\hline
 \multicolumn{3}{c}{Lower level} & & \multicolumn{3}{c}{Upper level} & Rel. Int. \\
 \cline{1-3} \cline{5-7} \\
       $J,L,S$    & $g$ & $m$        & & $J,L,S$           & $g$         & $m$            & \\
\hline
                      \multicolumn{8}{c}{NaI\,$\lambda$5889.951} \\
$1/2,0,1/2$   & 2     & $-1/2$   & & $3/2,1,1/2$   & $4/3$         & $-3/2$         &  $3/16$ \\
              &       & $-1/2$   & &               &               & $-1/2$         &  $1/4$ \\
              &       & $1/2$    & &               &               & $-1/2$         &  $1/16$ \\ 
              &       & $-1/2$   & &               &               & $1/2$          &  $1/26$ \\
              &       & $1/2$    & &               &               & $1/2$          &  $1/4$ \\
              &       & $1/2$    & &               &               & $3/2$          &  $3/16$ \\
                      \multicolumn{8}{c}{NaI\,$\lambda$5895.924} \\
$1/2,0,1/2$   & 2     & $-1/2$   & & $1/2,1,-1/2$  & $4/3$         & $-1/2$         & $1/4$ \\
              &       & $1/2$    & &               &               & $-1/2$         & $1/4$ \\
              &       & $-1/2$   & &               &               & $1/2$          & $1/4$ \\
              &       & $1/2$    & &               &               & $1/2$          & $1/4$ \\
                      \multicolumn{8}{c}{MgI\,$\lambda$3832.304} \\
$1,1,0$       & $3/2$ & $-1$     & & $2,2,0$       & $4/3$         & $-2$           & $3/20$ \\
              &       & $-1$     & &               &               & $-1$           & $3/20$ \\
              &       & $0$      & &               &               & $-1$           & $3/40$ \\
              &       & $-1$     & &               &               & $0$            & $1/40$ \\
              &       & $0$      & &               &               & $0$            & $1/5$ \\
              &       & $1$      & &               &               & $0$            & $1/40$ \\
              &       & $0$      & &               &               & $1$            & $3/40$ \\
              &       & $1$      & &               &               & $1$            & $3/20$ \\
              &       & $1$      & &               &               & $2$            & $3/20$ \\
\hline
\end{tabular}
\end{table}
\addtocounter{table}{-1}
\begin{table}[h!]
\caption{Continued. L\'ande factors and relative intensities of the most detectable metals.}
\label{tbl-split2}
\centering
\begin{tabular}{ccrcccrc}
\hline\hline
 \multicolumn{3}{c}{Lower level} & & \multicolumn{3}{c}{Upper level} & Rel. Int. \\
 \cline{1-3} \cline{5-7} \\
       $J,L,S$    & $g$ & $m$        & & $J,L,S$           & $g$         & $m$            & \\
\hline
                      \multicolumn{8}{c}{MgI\,$\lambda$3838.292} \\
$2,1,1$       & $5/3$ & $-2$     & & $3,2,1$       & $4/3$         & $-3$           & $3/28$ \\
              &       & $-2$     & &               &               & $-2$           & $1/14$ \\
              &       & $-1$     & &               &               & $-2$           & $1/14$ \\
              &       & $-2$     & &               &               & $-1$           & $1/140$ \\
              &       & $-1$     & &               &               & $-1$           & $4/35$ \\
              &       & $0$      & &               &               & $-1$           & $3/70$ \\
              &       & $-1$     & &               &               & $0$            & $3/140$ \\
              &       & $0$      & &               &               & $0$            & $9/70$ \\
              &       & $1$      & &               &               & $0$            & $3/140$ \\     
              &       & $0$      & &               &               & $1$            & $3/70$ \\
              &       & $1$      & &               &               & $1$            & $4/35$ \\
              &       & $2$      & &               &               & $1$            & $1/140$ \\
              &       & $1$      & &               &               & $2$            & $1/14$ \\
              &       & $2$      & &               &               & $2$            & $1/14$ \\
              &       & $2$      & &               &               & $3$            & $3/28$ \\
                      \multicolumn{8}{c}{AlI\,$\lambda$3944.006} \\
$1/2,1,-1/2$  & $2/3$ & $-1/2$   & & $1/2,0,1/2$   & 2             & $-1/2$         & $1/4$ \\
              & $2/3$ & $1/2$    & &               &               & $-1/2$         & $1/4$ \\
              & $2/3$ & $-1/2$   & &               &               & $1/2$          & $1/4$ \\
              & $2/3$ & $1/2$    & &               &               & $1/2$          & $1/4$ \\
                      \multicolumn{8}{c}{AlI\,$\lambda$3961.520} \\
$3/2,1,1/2$   & 4/3 & $-3/2$     & & $1/2,0,1/2$   & 2             & $-1/2$         &  $3/16$ \\
              &     & $-1/2$     & &               &               & $-1/2$         &  $1/4$  \\
              &     & $1/2$      & &               &               & $-1/2$         &  $1/16$ \\
              &     & $-1/2$     & &               &               & $1/2$          &  $1/16$ \\
              &     & $1/2$      & &               &               & $1/2$          &  $3/16$ \\
              \hline
\end{tabular}
\end{table}
\addtocounter{table}{-1}
\begin{table}[h!]
\caption{Continued. L\'ande factors and relative intensities of the most detectable metals.}
\label{tbl-split3}
\centering
\begin{tabular}{ccrcccrc}
\hline\hline
 \multicolumn{3}{c}{Lower level} & & \multicolumn{3}{c}{Upper level} & Rel. Int. \\
 \cline{1-3} \cline{5-7} \\
       $J,L,S$    & $g$ & $m$        & & $J,L,S$           & $g$         & $m$            & \\
\hline
                      \multicolumn{8}{c}{CaII\,$\lambda$3933.663} \\
$1/2,0,1/2$   & 2 & $-1/2$       & & $3/2,1,1/2$   & $4/3$         & $1/2$          & $1/16$ \\
              &   & $1/2$        & &               &               & $3/2$          & $3/16$ \\
              &   & $-1/2$       & &               &               & $-1/2$         & $1/4$  \\
              &   & $1/2$        & &               &               & $1/2$          & $1/4$  \\
              &   & $-1/2$       & &               &               & $-3/2$         & $3/16$ \\
              &   & $1/2$        & &               &               & $-1/2$         & $1/16$ \\
\hline
           \multicolumn{8}{c}{CaII\,$\lambda$3968.469} \\
$1/2,0,1/2$   & 2 & $-1/2$       & & $1/2,1,1/2$   & $2/3$         & $1/2$          & $1/4$  \\
              &   & $-1/2$       & &               &               & $-1/2$         & $1/4$  \\
              &   & $1/2$        & &               &               & $1/2$          & $1/4$  \\
              &   & $1/2$        & &               &               & $-1/2$         & $1/4$  \\
\hline
           \multicolumn{8}{c}{CaI\,$\lambda$4226.728} \\
$0,0,0$       & 0 & $0$          & & $1,1,0$       & $1$           & $1$            & $1/4$  \\
              &   & $0$          & &               &               & $0$            & $1/2$  \\
              &   & $0$          & &               &               & $-1$           & $1/4$  \\
\hline
\end{tabular}
\end{table}

\end{document}